\documentclass[%
 reprint,showpacs,
 amsmath,amssymb,
 aps,
prd,
]{revtex4-1}

\usepackage{graphicx,color}
\usepackage{dcolumn}
\usepackage{multirow}
\usepackage{bm}

\begin{document}


\title{Tidal Deformabilities and Neutron Star Mergers}

\author{Tianqi Zhao}
\email{tianqi.zhao@stonybrook.edu}
\affiliation{
Department of Physics \& Astronomy, Stony Brook University, Stony Brook, NY 11794-3800}
\author{James M. Lattimer}
\email{james.lattimer@stonybrook.edu}
\affiliation{
Department of Physics \& Astronomy, Stony Brook University, Stony Brook, NY 11794-3800}

\date{\today}

\begin{abstract}
  Finite size effects in a neutron star merger are manifested, at
  leading order, through the tidal deformabilities of the stars.  If
  strong first-order phase transitions do not exist within neutron
  stars, both neutron stars are described by the same equation of 
  state, and their tidal deformabilities are highly correlated through
  their masses even if the equation of state is unknown.  If, however, a strong phase transition exists between
  the central densities of the two stars, so that the more massive star has a phase transition and the least massive star does not, this correlation will be
  weakened.  In all cases, a minimum deformability for each neutron star mass is imposed by causality, and a less conservative limit is imposed by the unitary gas constraint, both of which we compute.  In order to make the best use of gravitational wave data
  from mergers, it is important to include the correlations relating
  the deformabilities and the masses as well as lower limits to the deformabilities as a function of mass.  Focusing on the case without
  strong phase transitions, and for mergers where the chirp mass
  ${\cal M}\le1.4M_\odot$, which is the case for all observed double
  neutron star systems where a total mass has been accurately
  measured, we show that the ratio of the dimensionless tidal deformabilities
  satisfy $\Lambda_1/\Lambda_2\sim q^6$, where $q=M_2/M_1$
  is the binary mass ratio; $\Lambda$ and $M$ are the dimensionless deformability and mass of each star, respectively.  Moreover, they are bounded by $q^{n_-}\ge\Lambda_1/\Lambda_2\ge q^{n_{0+}+qn_{1+}}$, where
  $n_-<n_{0+}+qn_{1+}$; the parameters depend only on ${\cal M}$, which is
  accurately determined from the gravitational-wave signal. We also
  provide analytic expressions for the wider bounds that exist in the
  case of a strong phase transition.  We argue that bounded ranges for
  $\Lambda_1/\Lambda_2$, tuned to ${\cal M}$, together with lower bounds to $\Lambda(M)$, will be more useful in
  gravitational waveform modeling than other suggested approaches.

\end{abstract}
\pacs{95.85.Sz, 26.60.Kp, 97.80.-d}
\maketitle

\section{Introduction}
Finite size effects in a binary neutron star merger are manifested, to
lowest order, through the tidal deformabilities of the individual
stars.  The tidal effects are imprinted in the gravitational-wave
signal through the binary tidal deformability~\citep{Flanagan08,Hinderer08}
\begin{equation}
\tilde\Lambda={16\over13}{(12q+1)\Lambda_1+(12+q)q^4\Lambda_2\over(1+q)^5},
\label{eq:tildelambda}
\end{equation}
where $q=M_2/M_1\le1$ is the binary mass ratio.  The dimensionless deformability of each star is
\begin{equation}
\Lambda_{[1,2]}={2\over3}k_{2,[1,2]}\left({R_{[1,2]}c^2\over GM_{[1,2]}}\right)^5,
\label{eq:lambda}
\end{equation}
where $k_2$ is the tidal Love number~\citep{Damour09,Flanagan08,Hinderer08}, which is the
proportionality constant between an external tidal field and the
quadrupole deformation of a star.  $R_{[1,2]}$ and $M_{[1,2]}$
are the radii and masses of the binary components, respectively. $k_2$
can be readily determined from a first-order differential equation
simultaneously integrated with the two usual TOV structural equations~\citep{Hinderer10,Postnikov10} and has values ranging from about 0.05
to 0.15 for neutron stars.  For black holes, $k_2=0$.  The tidal
deformations of the neutron stars result in excess dissipation of
orbital energy and speed up the final stages of the inspiral.  Tidal
deformations act oppositely to spin effects, which tend to be more
important during earlier stages of the observed gravitational wave
signal.

The gravitational waves from the recently observed merger of two
neutron stars, GW170817, were analyzed by the LIGO/VIRGO
collaboration~\citep{Abbott17} (hereafter LVC), and subsequently reanalyzed by De et al. \cite{De18} (hereafter DFLB$^3$) and also the LIGO/VIRGO
collaboration~\citep{Abbott18} (Hereafter LVC2).  In the LVC analysis, the
gravitational-wave signal was fitted to the Taylor F2 post-Newtonian
aligned-spin model~\citep{Sathyaprakash91,Buonanno09,Arun09,Mikoczi05,Bohe13,Vines11} which has 13 parameters.  7 of those
parameters are extrinsic, including the sky location, the source's
distance, polarization angle and inclination, and the coalescence
phase and time.  The remaining 6 parameters are intrinsic, including
the masses $M_1$ and $M_2$, dimensionless tidal deformabilities
$\Lambda_{[1,2]}$, and the component's aligned spins
$\chi_{[1,2]}=cJ_{[1,2]}/GM_{[1,2]}^2$, where $J$ is the angular
momentum.  The reanalysis of DFLB$^3$ differed from that of LVC
chiefly in that electromagnetic observations were used to fix the
source location and distance and in the adoption of the relation
$\Lambda_1/\Lambda_2=q^6$, expressing the assumption that the two stars have a common equation of state (EOS).
They justified this assumption using parameterized hadronic EOSs
modeled using a fixed neutron star crust and three high-density
polytropic segments whose parameters were restricted by causality and
a minimum value of an assumed neutron star maximum mass.  DFLB$^3$ also employed
the causal lower limit to $\Lambda(M)$ in their analysis.  In contrast,
the analysis of LVC assumed uncorrelated priors for $\Lambda_1$ and
$\Lambda_2$, thereby assuming that the two stars did not have the same
equation of state, and did not consider causality-violating values of $\Lambda_1$ or $\Lambda_2$.  DFLB$^3$ showed that models including correlations were favored by odds ratio $\gtrsim100$ over models
using uncorrelated deformabilities, and, furthermore, that including
deformability correlations reduced the 90\% confidence upper limit to
the binary deformability by about 20\%.  The latter result was
confirmed by LVC2, who reanalyzed the GW170817 signal including
deformability correlations using two different prescriptions.

It is reasonable to assume that future investigations of neutron star
mergers will treat $\Lambda_1$ and $\Lambda_2$ as correlated
parameters, irrespective of which waveform model is used. The purposes
of this paper are 1) to replace the approximate result
$\Lambda_1/\Lambda_2=q^6$ with analytic bounds suitable for
use in existing methods of fitting gravitational-wave signals of
neutron star mergers, 2) to establish realistic lower limits to $\Lambda(M)$, 3) to compare our method with one proposed by
Yagi and Yunes \cite{Yagi17}, and 4) to determine modifications to deformability
correlations due to the possible existence of a strong first order phase
transitions in the density range between the central densities of the
two stars.  In this case, the more massive star will be considered to
be a {\it hybrid} star, in contrast to the lower mass star which we
refer to as a {\it hadronic} star.  This oversimplified notation harks
back to the possibility of a hybrid hadronic-quark matter star in
which the quark matter-hadronic matter interface has a surface tension
too large to permit a smooth Gibbs phase transition.  In the event of
a strong first order phase transition, the more massive star can have
a radius and tidal deformability much smaller than the lower mass
star, even though their masses are nearly equal.  This weakens the
correlations otherwise evident between the tidal deformabilities and
masses.

In addition to bounds on the deformability ratio
$\Lambda_1/\Lambda_2$, future analyses will benefit from the
incorporation of absolute lower bounds to $\Lambda(M)$ available from
consideration of the {\it maximally compact} EOS~\citep{Koranda97,Lattimer12}, which
are limited by causality and the observed minimum value of the neutron
star maximum mass.  This EOS assumes that the matter pressure is
essentially zero below a fiducial density $n_o$ that is a few times
the nuclear saturation density, and that above this density the sound
speed is equal to the speed of light.  However, we also determine a
more realistic and less extreme lower bound in which the pressure in
the vicinity of the nuclear saturation density is instead limited from
below by the unitary gas constraint thought to be applicable for
neutron star matter~\citep{Tews17}.  Upper bounds to $\Lambda(M)$ are
available from nuclear theory and experiment, but are unfortunately
model-dependent, and astrophysical observations also cannot yet
provide accurate upper bounds.  We will, however, explore the
sensitivity of both lower and upper deformability bounds to
assumptions concerning the minimum pressure of neutron star matter and
also the minimum and maximum values assumed for the neutron star
maximum mass.

This paper is organized as follows:  \S \ref{sec:likely} describes the most likely masses and spins for merging neutron star systems, and \S \ref{sec:tide} reviews how tidal deformabilites are defined and calculated.  \S \ref{sec:par} outlines the parameterized equations of state used in this paper and the resulting tidal deformabilities and their bounds, while \S \ref{sec:bin} outlines results for the binary tidal deformabilities and their bounds. \S \ref{sec:hadron} establishes the correlations of tidal deformabilities with masses and compares our approach with other work.  The lower bounds on deformabilities from causality are summarized in \S \ref{sec:caus}, and those from the unitary gas and neutron matter constraints are discussed in \S \ref{sec:unitary}.  Deformability constraints for hybrid stars are established in \S \ref{sec:hybrid}.  We summarize our conclusions in \S \ref{sec:conclusion}.

\section{Likely Mass and Spin Ranges for Observable Merging Neutron Star Systems\label{sec:likely}}
\begin{figure*}
\hspace{-0.4cm}\includegraphics[width=0.5\linewidth,angle=180]{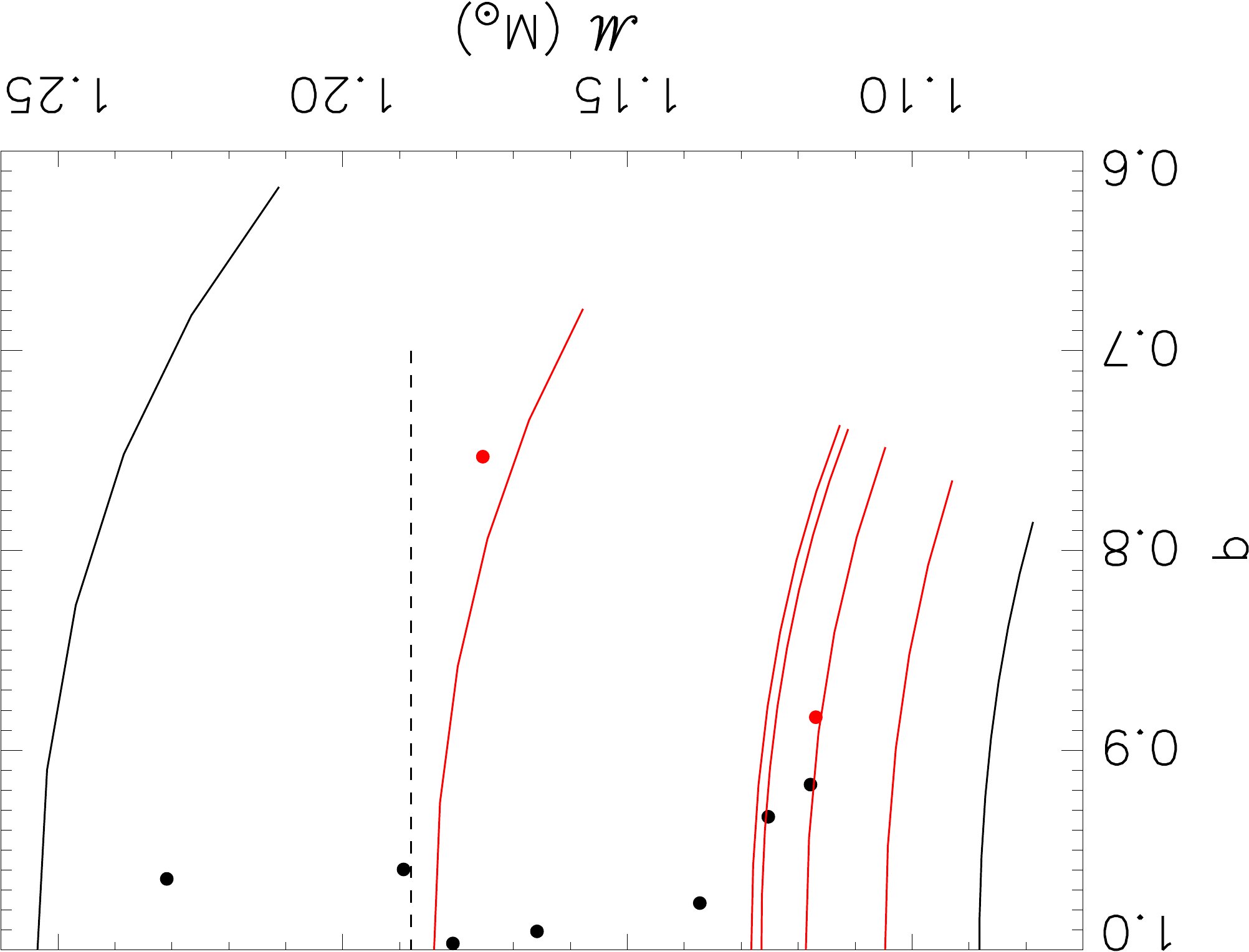}
\hspace{-0cm}\includegraphics[width=0.512\linewidth,angle=180]{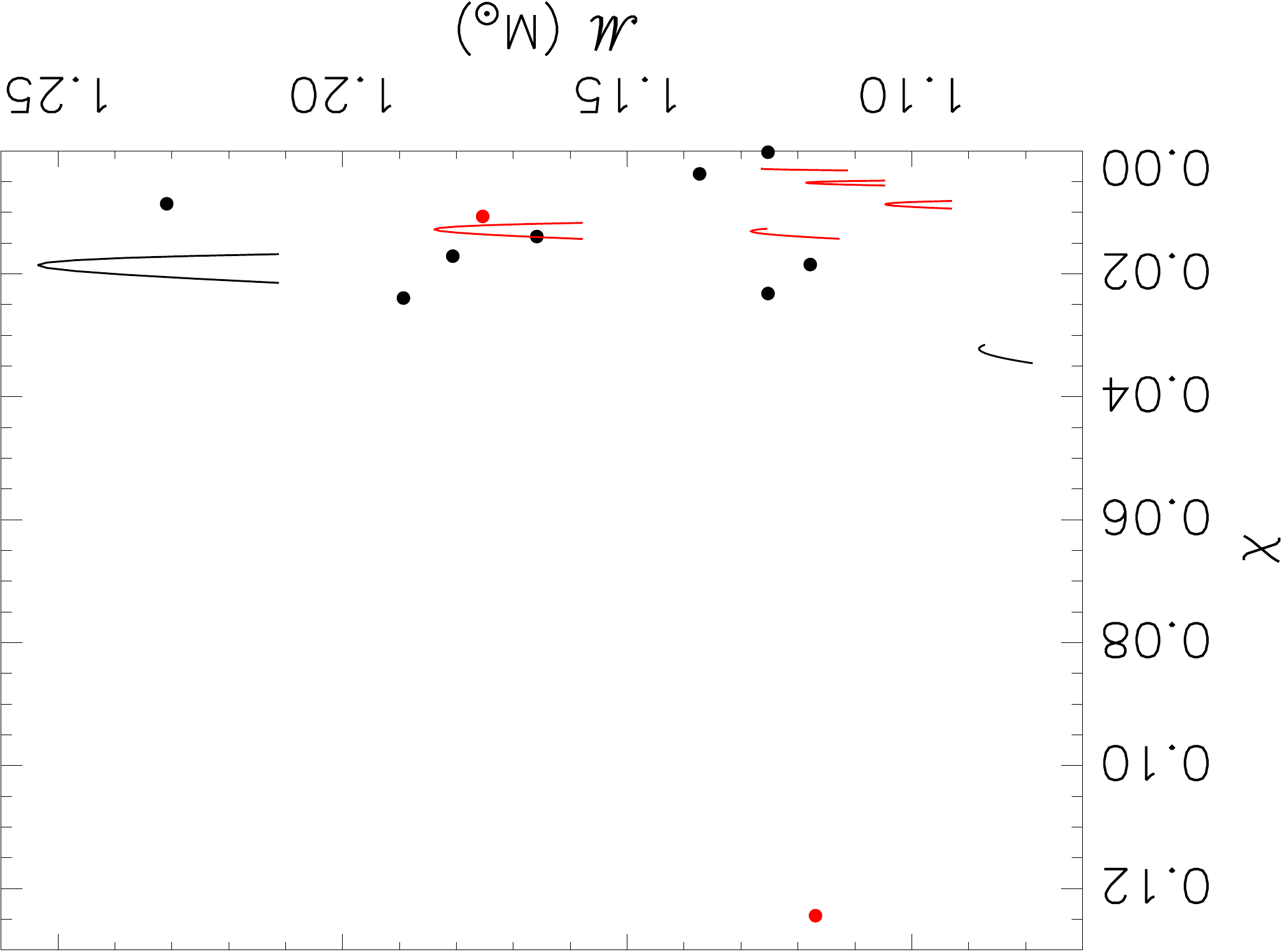}
\caption{Left: Binary mass ratio $q$ as a function of chirp mass
  ${\cal M}$ for known double neutron star (DNS) systems~\citep{Tauris17}. ${\cal M}$ for GW170817 is indicated by the vertical dashed line.  Right: Spin parameters for pulsars in known DNS systems.   For both figures,   curves
  represent possible values for systems in which the total mass,
  but not $q$, is accurately known; the minimum value of $q$ is
  determined by $M_2>1.1M_\odot$. Red curves and points indicate
  systems for which the merger timescale $\tau_{GW}$ is longer than the age of the
  Universe.\label{fig:DNS}}
\end{figure*}

It seems likely that future observations of merging neutron stars,
like GW170817, will have component masses and spins similar to those
of known double neutron star systems (DNS).  Known systems contain at
least one pulsar and their masses and spins have been determined by
pulsar timing.  There are 9 systems in which both masses are
accurately determined, and 7 others for which only the total mass
$M_T=M_1+M_2$ is known with precision~\citep{Tauris17}.  Determination of $q$ and
${\cal M}$ for the former systems is straightforward.  However, even
in the latter cases, some information about ${\cal M}$ and $q$ can be
established, using the theoretical paradigm that the minimum neutron star mass is $\gtrsim1.1M_\odot$ (for further discussion, see Ref. \cite{Lattimer12}).
Note that we can write
\begin{eqnarray}
\qquad {\cal M}&=&{M_1^{3/5}M_2^{3/5}\over M_T^{1/5}}=M_T^{2/5}\left(1-{M_2\over M_T}\right)^{3/5}M_2^{3/5},\cr
q&=&{M_2\over M_1}={M_2\over M_T-M_2}
\label{eq:m2}
\end{eqnarray}
so the restriction $1.1M_\odot\le M_2<M_T/2$ determines ${\cal M}(q)$. Values for ${\cal M}$ and $q$ for known DNS are shown in Fig. \ref{fig:DNS}.   Two systems have $q<0.9$, but also have gravitational decay times $\tau_{GW}$ longer than the age of the universe and so may not be representative of observed merging systems. 

In the same way, the spin parameters $\chi=2\pi Ic/(GM^2P)$ of the
pulsars in these systems, where $P$ is the spin period and $I$ the
moment of inertia, can be estimated. One system, PSR 3039-0737~\cite{Ferdman13},
contains two pulsars, so there are 10 pulsars with known masses and
spins in these systems. Using the piecewise polytrope ansatz
(see below) in the slow-rotation limit, it has been determined~\citep{Steiner16} that
\begin{equation}
{I\over MR^2}\simeq 0.01+1.2\beta^\frac{1}{2}-0.1839\beta-3.735\beta^{3\over2}+5.278\beta^2,
\label{eq:i}
\end{equation}
where $\beta=GM/Rc^2$ is the compactness parameter and $R$ is the circumferential stellar radius, 
assuming that the minimum neutron star maximum mass is $1.97M_\odot$.
Using $R\simeq12$ km, estimates for $\chi$ are also displayed in
Fig. \ref{fig:DNS}.  These estimates do not reflect the fact that the
spins at merger in almost all systems will be much smaller than their
current values.  For example, PSR 1913+16A, with ${\cal
  M}=1.23M_\odot$, has $\tau_{GW}\dot P\simeq1.3P$~\citep{Weisberg10}.
Note that one star (J1807--2500B, which might not even be a DNS
system~\citep{Lynch12}) has $\chi\simeq0.12$, much larger than the
other 15 cases, but exists in a system with $\tau_{GW}$ longer
than the Universe's age and so may not be typical of an observed
merging system.

It therefore seems reasonable to assume that potential future mergers,
like GW170817, will have $1M_\odot\le{\cal M}\le1.3M_\odot$, $0.9\le
q\le1$ and component spin parameters $\chi\lesssim0.02$.  Calculation of the tidal deformabiities and moments of inertia in the
slow-rotation limit seems justified.

\section{Calculation of Tidal Deformabilities\label{sec:tide}}

The dimensionless tidal deformabiity parameter $\Lambda$ can be calculated in the small quadrupole deformation limit from~\citep{Hinderer10}
\begin{multline}
   \Lambda={16g\over15}[4\beta^2(3-9\beta+4\beta^2+6\beta^3)+3g\ln(1-2\beta)\\
   -2\beta z_R(1-\beta)(1-2\beta)(3-6\beta-2\beta^2)]^{-1}
\label{eq:lamb} 
\end{multline}

where 
\begin{equation}
 g=[2\beta(1+z_R)-z_R]\left(1-\beta\right)^2.
\end{equation}
$z_R=z(R)$ is the surface value of the variable $z(r)$ determined by the first-order equation~\citep{Postnikov10}
\begin{equation}
{dz\over dr}= {f_1-f_2+f_3\over r(r-2Gm/c^2)}
\label{eq:z}
\end{equation}
with the boundary condition at the origin $z(r=0)=0$, and
\begin{eqnarray}
f_1&=&zr\left[\left(1-{2Gm\over rc^2}\right)(4+z)+1\right]-{8Gm\over c^2},\cr
f_2&=&{4G^2\over c^4}{(m+4\pi pr^3/c^2)^2\over r-2Gm/c^2},\cr
f_3&=&{4\pi G r^3\over c^4}\left[(2+z)(p-\varepsilon)+5\varepsilon+9p+{\varepsilon+p\over c_s^2/c^2}\right].
\label{eq:fff}
\end{eqnarray}
$m$, $p$ and $\varepsilon$ are the enclosed mass, pressure and mass-energy density at the radius $r$, respectively, related
by the usual general relativistic structure equations.
Note the appearance of the sound speed
$c_s=c\sqrt{\partial p/\partial\varepsilon}$ in Eq. (\ref{eq:fff}).  In
the case of a first-order phase transition in which a discontinuity
$\Delta\varepsilon_t$ occurs at the radius $r_t$ where the pressure
and enclosed mass are $p_t$ and $m_t$, respectively, and $c_s=0$
within the transition, a correction term
$\Delta z=-4\pi\Delta\varepsilon_tr_t^3/{m_t c^2}$~\citep{Postnikov10} must be added
to $z$ at the radius $r_t$.  In the case of small $\beta\lesssim 0.1$,
there are severe cancellations in Eq. (\ref{eq:lamb}), and a Taylor
expansion in $\beta$~\citep{Postnikov10} is utilized for accuracy.
However, we only consider neutron stars with $M\ge1.1M_\odot$ for which
$\beta\gtrsim 0.11$.

\section{Parameterized Equations of State and the Tidal Deformability\label{sec:par}}

The intrinsic parameters describing neutron stars in gravitational
waveform modeling include the component masses, spins and tidal
deformabilities. Spins are described by the dimensionless spin
parameters $\chi_1$ and $\chi_2$, while the deformabilities are
described by the parameters $\Lambda_1$ and $\Lambda_2$ for
nonspinning stars.  For nonspinning stars, $\Lambda$ is determined only by $M$ for a given EOS.  Even though the EOS is {\it a
  priori} unknown, it is nevertheless bounded by general considerations such as
thermodynamic stability, causality, the necessity to produce stars
with a minimum value of $M_{max}$, and nuclear physics considerations.  Therefore, values of
$\Lambda_2$ and $\Lambda_1$, for specified values of $m_1$ and $m_2$, must also
be bounded.  These bounds appear as correlations among $\Lambda_1,\Lambda_2,M_1$ and $M_2$.

In their analysis of GW170817, LVC did not take any correlations among
$\Lambda_1, \Lambda_2, M_1$ and $M_2$ into account.  DFLB$^3$, for reasons
summarized below, adopted the correlation $\Lambda_1/\Lambda_2=q^6$
and were able to show that models with this deformability correlation
were favored relative to models without it by odds ratio greater than
100.  Furthermore, they showed that including deformability
correlations generally reduced the 90\% confidence upper limit to the
binary deformability by about 20\% (a result confirmed by LVC2).
However, since the EOS is uncertain, the ratio $\Lambda_1/\Lambda_2$
has a finite range around the value $q^6$.  LVC2 used the methodology
of Ref.~\cite{Yagi17,Chatziioannou18} to estimate this range statistically
from fits to realistic EOSs.  Instead we will determine
bounds to $\Lambda_1/\Lambda_2$ in an EOS-insensitive
fashion, using causality and the observed minimum value for the
neutron star maximum mass.  We will compare this approach with that
adopted by LVC2 in \S \ref{sec:hadron}.

We will bound $\Lambda_1/\Lambda_2$ as a function of $q$ using
thousands of equations of state computed using the piecewise-polytrope
methodology~\cite{Read09,Ozel09,Steiner16,Lattimer16}. We find that these bounds can be
expressed in terms of particularly simple analytic forms.  Although Ref.~\cite{Lindblom18} argues that piecewise polytropes are less accurate
than other methods, such as spectral decomposition,
accuracy is not a consideration.  Rather, we are only interested in
the allowed range of deformabilities.  In fact, since the spectral
decomposition technique smooths equations of state near segment
boundaries, it actually misses some possibilities compared to
piecewise polytropes and may understate the true bounds.
The same is true for the
QCD-motivated scheme of Ref.~\cite{Kurkela14} which requires all EOSs to asymptotically approach $c_s=c/\sqrt{3}$ at high densities.

Read et al.~\cite{Read09} found that high-density cold equations of state could be relatively faithfully modeled with three polytropic segments coupled to a crust equation of state.  The crust equation of state applies for densities below  $n_0\sim n_s/2$, where $n_s=0.16$ fm$^{-3}$ is the nuclear saturation density; this region is dominated by nuclei in a Coulomb lattice together with a neutron liquid in chemical potential and pressure equilibrium.  The details of the crust equation of state are not important as differences among existing models produce very small effects for the structure of stars more massive than a solar mass. Each segment is described by the polytropic equation of state $p=K_in^{\gamma_i}$ for the region $n_{i-1}<n<n_i$ for $ i=1-3$ where $p$ is the pressure. Knowledge of $n_0$ and $p_0$, and continuity of $p$ and the energy density $\varepsilon$ at the boundaries, determines $K_i$ and leaves 6 free parameters, $n_i$ and $\gamma_i$ for $i=1-3$, or, equivalently, $n_i$ and $p_i$. Within the polytropic segment $i$, the energy density is given by
\begin{equation}
\varepsilon=\varepsilon_{i-1}{n\over n_{i-1}}+{p-p_{i-1}(n/n_{i-1})\over\gamma_i-1},\quad n_{i-1}\le n\le n_i.
\end{equation}
The polytropic indices and the energy densities at the boundaries are given by
\begin{eqnarray}
\quad\varepsilon_i&=&{p_i\over\gamma_i-1}+\left(\varepsilon_{i-1}-{p_{i-1}\over\gamma_i-1}\right){n_i\over n_{i-1}}, \cr
\gamma_i&=&{\ln(p_i/p_{i-1})\over\ln(n_i/n_{i-1})} \qquad i=1,2,3.\label{eq:ebound}
\end{eqnarray}

Ref. \cite{Read09} made the additional observation that a wide variety of equations of state could be accurately described with a single set of boundary densities: $n_3\simeq2n_2\simeq4n_1\simeq7.4n_s$.  Assuming these values leaves three free parameters $p_i$ for $i=1-3$.  We stress that a specific equation of state could be more accurately modeled with a larger number of segments, but we are chiefly concerned with achieving an exhaustive coverage of pressure-energy density (or mass-radius) space.  We have shown that adding more segments does not expand this coverage significantly for hadronic stars.  In \S\ref{sec:hybrid}, we add additional parameters to ensure a complete coverage of the possibility of hybrid configurations.  

Some results for neutron star structure with the piecewise polytrope methodology have been previously reported~\citep{Steiner16,Lattimer16}.  We summarize here our specific assumptions: 
\begin{itemize}
\item Neutron stars have hadronic crusts which terminate at the
fixed  density $n_0=n_s/2.7$, where $p_0=0.2177$ MeV fm$^{-3}$,  $\varepsilon_0=56.24$ MeV fm$^{-3}$ and $e_0=\varepsilon_0/n_0-mc^2=9.484$ MeV, values obtained by interpolating the SLy4 EOS~\cite{Chabanat98}. Here, $e(n,x)$ is the internal energy per baryon and $mc^2=939.566$ MeV.
\item The first polytropic segment between $n_0$ and
  $n_1=1.85n_s$ is constrained by neutron matter calculations~\cite{Drischler16} such that
  8.4 MeV fm$^{-3}\lesssim p_1\lesssim 20$ MeV fm$^{-3}$ used in our previous studies.  However, we deliberately choose here a 50\% larger upper bound, 30 MeV fm$^{-3}$, in order to obtain values of $\tilde\Lambda$ that are well above the 90\% confidence limit inferred from the LVC analysis of GW170817.  We also consider a smaller lower limit to $p_1$, 3.74 MeV fm$^{-3}$, arising from the unitary gas constraint~\cite{Tews17}, separately in \S \ref{sec:unitary}.   We note the value of $p_1$ effectively determines the nuclear symmetry energy $S_v$ and its slope parameter $L$ at the nuclear saturation density.  Assuming that higher-than-quadratic terms in the Taylor expansion of the nuclear energy per particle $e(n,x)$ in powers of the neutron excess $1-2x$ are negligible near $n=n_s$, and also that the proton fraction $x\sim0$, one has
\begin{align}
S_V&=e(n_s,0)-e(n_s,1/2)\cr
&=e_0+B+{p_0\over n_0(\gamma_1-1)}\left[\left({n_s\over n_0}\right)^{\gamma_1-1}-1\right],\cr 
L&={3p(n_s,0)\over n_s}=3{p_0\over n_0}\left({n_s\over n_0}\right)^{\gamma_1-1},
\label{eq:svl}\end{align}
where $B=-e(n,1/2)\simeq16$ MeV is the bulk binding energy of symmetric matter.  We find using Eqs. (\ref{eq:ebound}) and (\ref{eq:svl}) that $2.27\le\gamma_1\le3.06$,  33.4 MeV $<S_V<37.5$ MeV and 38.9 MeV $<L<85.3$ MeV, approximately the ranges predicted by nuclear experiments and neutron matter theoretical calculations~\cite{Lattimer13}, except for $S_V$ which is about 2 MeV larger due to the polytropic approximation.

\begin{figure*}[ht]
\includegraphics[width=\linewidth,angle=0]{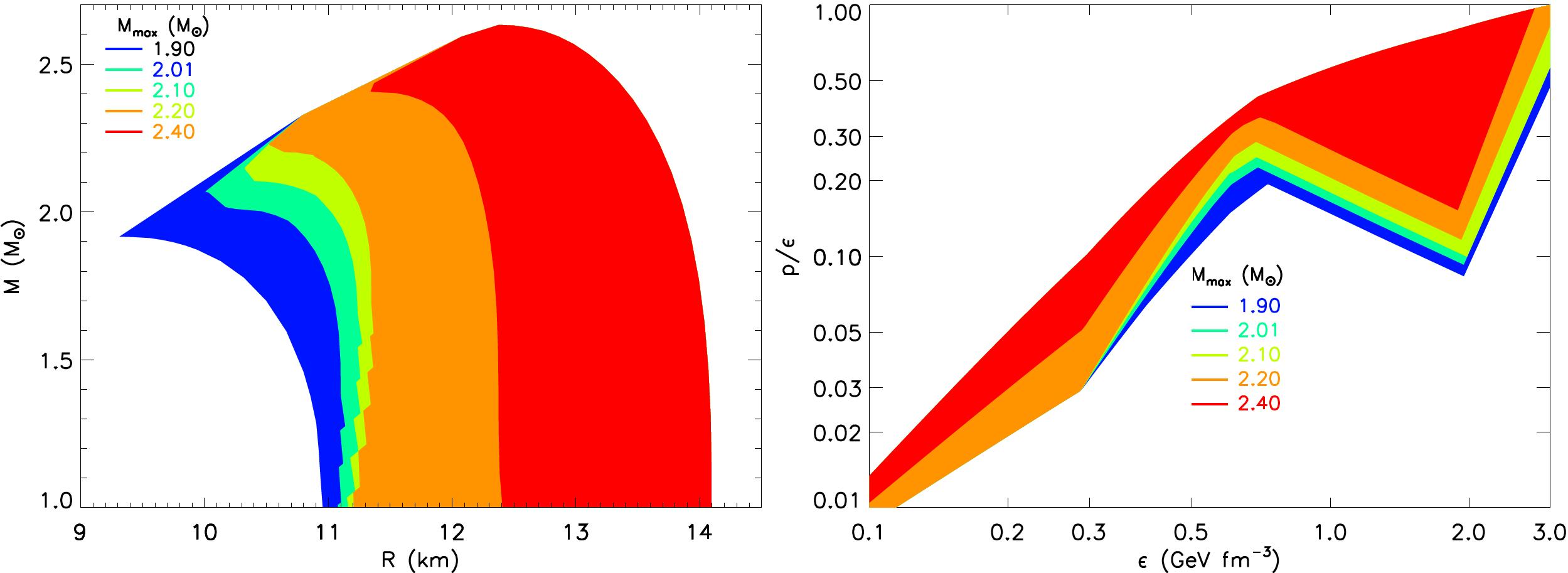}
\caption{Left panel:  Permitted values of masses and radii for different assumptions about the minimum neutron star maximum mass $M_{max}$  A minimum value of $p_1=8.4$ MeV fm$^{-3}$ was assumed.  Right panel: Permitted values of pressure and energy density for different assumptions about $M_{max}$.\label{fig:mvsr}}
\end{figure*}

\item The parameter $p_2$ is limited from above by enforcing causality ($c_s^2/c^2=\partial p/\partial\varepsilon\le1$) at $n_2$,
 which results in the implicit equation for the upper bound to $\gamma_2$,
\begin{align}
\gamma_{2,max}&(\gamma_{2,max}-2)=\cr
&\left[(\gamma_{2,max}-1){\varepsilon_1\over p_1}-1\right]\left({n_1\over n_2}\right)^{\gamma_{2,max}-1}.
\end{align}
\item The parameter $p_3$ is limited from above by the condition $\gamma_{3,max}=1+\varepsilon_2/p_2$.  This value guarantees that causality is violated for the maximum mass configuration for any $p_1$ and $p_2$, but only configurations with $\gamma_3<\gamma_{3,max}$ (and thus $p_3<p_{3,max}$) that don't violate causality are ultimately accepted.
\item The parameters $p_2$ and $p_3$ are limited from below either by $p_3\ge p_2\ge p_1$, which guarantees thermodynamic stability, or the requirement that the maximum mass exceeds a fiducial value.
\end{itemize}

The parameters $p_1, p_2$ and $\ln p_3$ are uniformly sampled within
their respective ranges.  The neutron star mass, radius and tidal
deformability are found from integration of the normal TOV
differential equations together with Eq. (\ref{eq:z}), in which it is
only necessary to specify $p(n)$ and $dp/d\varepsilon$ as functions of
$\varepsilon(n)$.  For each parameter set, we compute a series of 50
configurations assuming central pressures in the range
($3\cdot10^{-5}- 2\cdot10^{-3}$) km$^{-2}$.  Note that 1 Mev fm$^{-3}$
corresponding to $1.32375\cdot10^{-6}$ km$^{-2}$.  The lowest central
pressure results in stars with $M\sim0.5M_\odot$.  The largest central
pressure is always beyond the value which obtains in the lowest
assumed maximum mass configuration, $1.90M_\odot$.  (The central
pressure of the maximum mass star decreases with increasing maximum
mass values~\citep{Lattimer11}).  The differential equations are
solved using $\ln p$ as the independent variable with a variable
step-size 4th-5th order Runge-Kutta scheme.  In every case, the
surface pressure is set to $3\times10^{-13}$ km$^{-2}$.  The total mass,
moment of inertia and tidal deformability are insensitive to the
surface pressure, but  the radius is not, so we employ an analytic correction to compensate
for non-zero surface pressures  (these are at most 0.1
km in the lowest mass stars).

The value of the neutron star maximum mass plays an important role in
the allowed ranges of neutron star masses and radii, as well as in the
allowed values of $p_2$ and $p_3$ which constrain the equation of
state.  The left panel of Fig. \ref{fig:mvsr} displays allowed masses
and radii as a function of the assumed lower limit to the neutron star
maximum mass.  Clearly, larger minimum values of the neutron star
maximum mass prohibit smaller neutron star radii for every mass and
more severely constrain allowed trajectories of the $M-R$ relation.
Nevertheless, the minimum value of $p_1$, $p_{1,min}$ is an important
factor determining the minimum neutron star radius.  We found that if
$p_{1,min}$ is reduced to the unitary gas minimum, 3.74 MeV fm$^{-3}$,
radii of $1.4M_\odot$ stars as low as 10.4 km may be achieved for
$M_{max}=1.90M_\odot$.  The maximum neutron star radius is determined
by the maximum value of $p_1$, $p_{1,max}$  but not by the maximum mass, as the radius is insensitive to the high-density equation of state.  These results straightforwardly follow from the fact that the pressure in the density range $1-2n_s$, i.e., $p_1$, and $R_{1.4}$, the radii of $1.4M_\odot$ stars, are known to be highly correlated~\cite{Lattimer01}.

The right panel of Fig. \ref{fig:mvsr} shows allowed regions of $p$ as
a function of $\varepsilon$, which show greater restrictions as the
minimum value of the neutron star maximum mass is increased.  At lower
densities, $\varepsilon\lesssim 300$ MeV fm$^{-3}$ (which corresponds to
$p\le p_1$), the effect of the maximum mass is small until
$M_{max}\gtrsim 2.3M_\odot$.  Recall that the saturation density
$n_s\simeq0.16$ fm$^{-3}$ corresponds to $\varepsilon\simeq150$ MeV
fm$^{-3}$.  But for higher densities, the maximum mass constraint
becomes important for smaller values of $M_{max}$.

\begin{figure}
\includegraphics[height=\linewidth,angle=90]{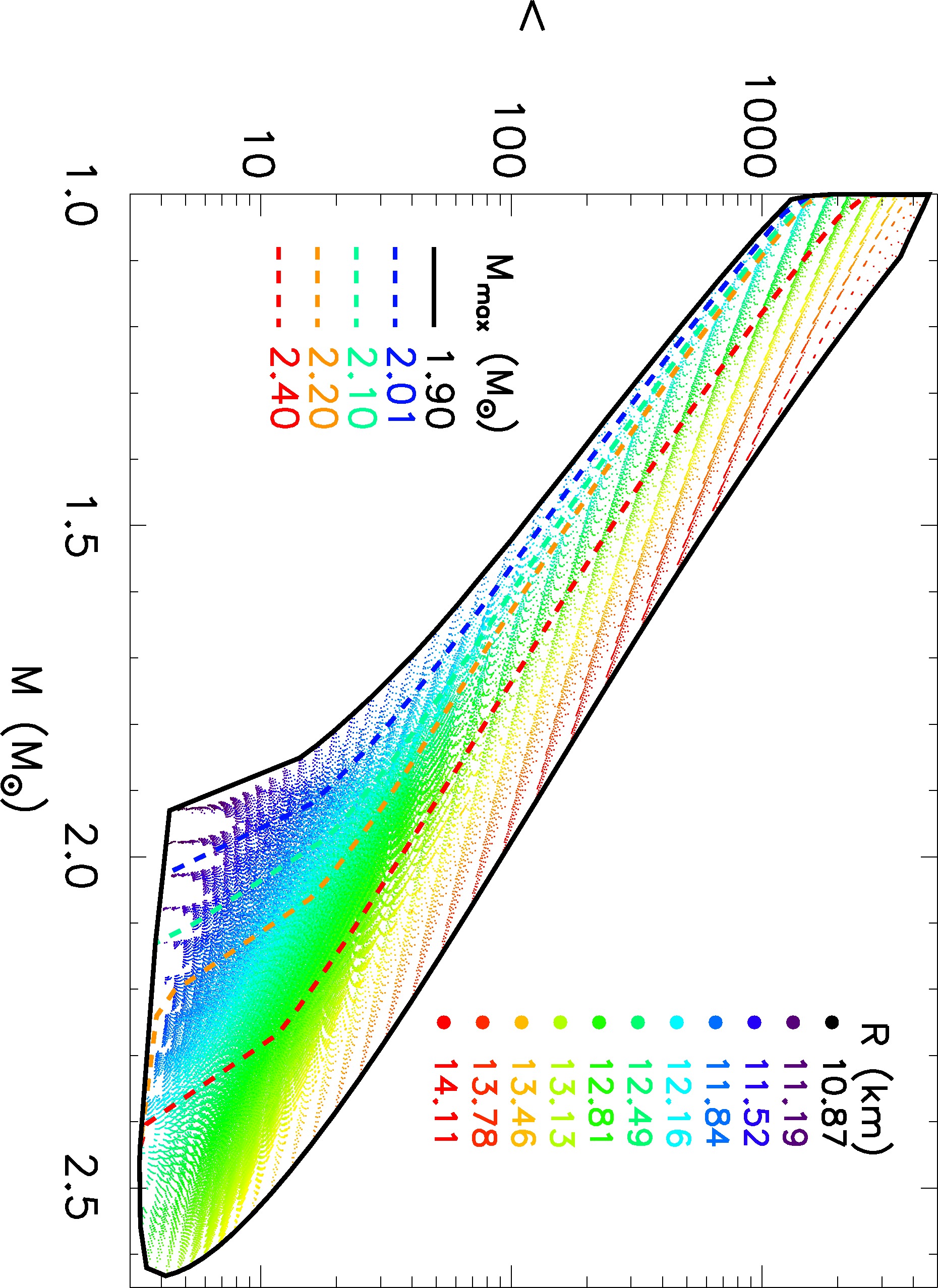}
\caption{The dimensionless tidal deformability for
  individual stars as a function of mass for various equations of state are marked by dots, which are color-coded by their radii.  Those configurations lying between the lower solid or colored dashed lines and the upper-most solid line originate from equations of state which
  satisfy the indicated $M_{max}$ constraint.  $p_{1,min}=8.4$ MeV fm$^{-3}$ and $p_{1,max}=30$ MeV fm$^{-3}$ were assumed.\label{fig:lam}}
\end{figure}
The dimensionless deformability as a function of $M$ and $R$ for causally-constrained piecewise polytropes are shown in Fig. \ref{fig:lam}.   In this figure,
individual configurations are color-coded according to their radii.  Clearly, there are well-defined upper and lower bounds for $\Lambda(M)$, with the upper (lower) bound defined by the stars with the largest (smallest) radii.  Thus, as found for radius bounds, the upper bound is determined by $p_{1,max}$ and is not sensitive to the assumed value of $M_{max}$ or $p_{1,min}$, while the lower bound is determined by both $p_{1,min}$ and  $M_{max}$.  The lower bound for $\Lambda(M)$ is an important constraint that should be taken into account in gravitational waveform modeling of BNS mergers, and is further explored in \S \ref{sec:unitary}. 

The fact that $\Lambda$ decreases rapidly with $M$ and increases rapidly with $R$ is not surprising given the formula $\Lambda=(2k_2/3)\beta^{-5}$.  However, we find for moderate masses that $\Lambda\propto\beta^{-6}$ provides a better description.  This follows because the
behavior $k_2\propto\beta^{-1}$ is observed ~\citep{Postnikov10,Hinderer10} for a wide variety of equations of state in the mass
range $1.1M_\odot\lesssim M\lesssim1.6M_\odot$ (corresponding to, roughly,
$0.11\lesssim\beta\lesssim0.20$).  This mass range is precisely
the range expected if observed double neutron star binaries are
typical merger candidates, and is the range of neutron star masses inferred for GW170817~\cite{Abbott17,De18}. Ref. \cite{Lattimer01} found that $R_{1.4}\propto p^{1/4}$ in the density range $n_s-2n_s$.  Given that $\Lambda\propto R^6$ for a given mass, and $n_s<n_1<2n_s$, it follows that $\Lambda_{1.4,max}\propto p_{1,max}^{5/4-3/2}$, which we find to approximately be the case.

\begin{figure}
\includegraphics[width=\linewidth,angle=180]{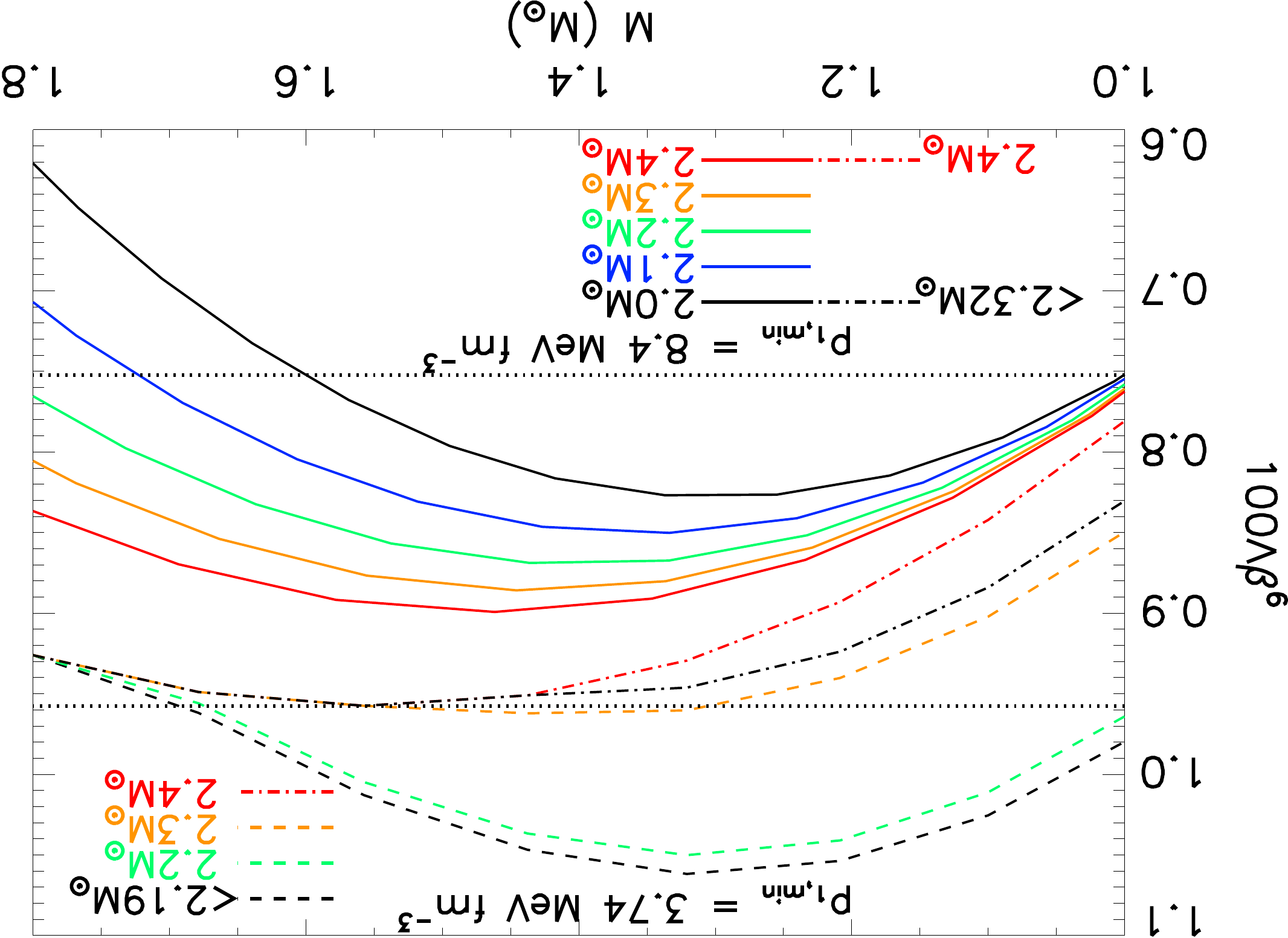}
\caption{Bounds of 
  $\Lambda\beta^6$ as a function of mass for piecewise polytropes as constrained  by $M_{max}$ and $p_{1,min}$.  $p_{1,max}=30$ MeV fm$^{-3}$ is assumed. Solid curves are lower bounds for the indicated $M_{max}$.  Upper bounds for $p_{1,min}=3.74~(8.4)$ MeV fm$^{-3}$ are shown by dashed (dot-dashed) curves. Note that for $1.1M_\odot<M<1.6M_\odot$ and $M_{max}>2M_\odot$ that $\Lambda\beta^6$ is constant to about $\pm12\%$ (dotted lines). \label{fig:lamb6}}
\end{figure}
These results are illustrated in
Fig. \ref{fig:lamb6}, which shows $\Lambda\beta^6$ as a function of
$M$.  We infer the important result that, in our relevant mass range,
\begin{equation}
\Lambda=a\beta^{-6},
\label{eq:lam6}
\end{equation}
where $a=0.0085\pm0.0010$ bounds the
results as long as $M_{max}\gtrsim2M_\odot$ and $p_{1,min}=8.4$ MeV
fm$^{-3}$.  Because $\Lambda$ is largely proportional to $R^6$, we find that, in contrast to the situation for $\Lambda$, the upper limit of $\Lambda\beta^6$ is insensitive to the value of $p_{1,max}$, and the lower limit is insensitive to $p_{1,min}$.  Nevertheless, the upper limit acquires a sensitivity to $p_{1,min}$ because $\Lambda\propto R^6$ is only approximate.  It is also noted that for $p_{1,min}=8.4$ MeV fm$^{-3}$ and $M_{max}\gtrsim2.32M_\odot$, or $p_{1,min}=3.74$ MeV fm$^{-3}$ and $M_{max}\gtrsim2.19M_\odot$, the upper boundary also depends on $M_{max}$.

We find that the upper bounds for both $\Lambda$ and $\Lambda\beta^6$ can be further reduced if one can impose an upper limit to the neutron star maximum mass,  as perhaps can be inferred for GW170817~\cite{Margalit17,Shibata17}.  However, these reductions are realized only for $M/M_{max}>0.75$, generally outside the interesting range for observed double neutron star binaries. The reductions increase as $M/M_{max}$ increases.  For $M_{max}\le2.2M_\odot$ and $M=M_{max}$, $\Lambda$ can be reduced by a factor of 2 and $\Lambda\beta^6$ can be reduced by about 0.001.  There is no change to the lower bound of either quantity. 
\section{The Binary Deformability\label{sec:bin}}
\begin{figure}
\includegraphics[height=\linewidth,angle=90]{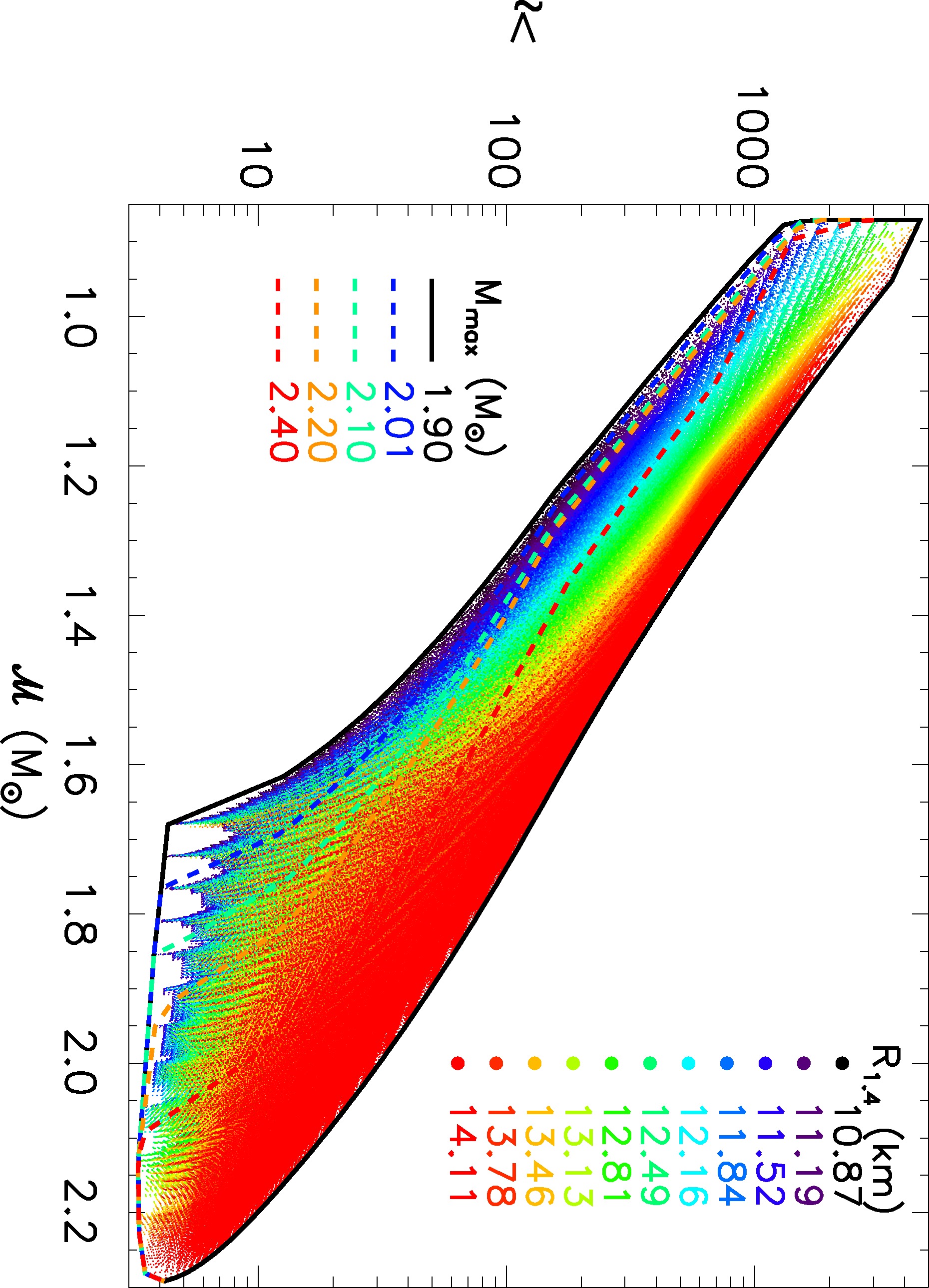}
\caption{Similar to Fig. \ref{fig:lam}, except that the dimensionless binary tidal deformability as
  a function of chirp mass is displayed, with stellar pairs  indicated with dots colored according to the value of $R_{1.4}$ for each assumed equation of state.  $M_{max}$ only affects the lower bound.  $p_{1,min}=8.4$ MeV fm$^{-3}$ is assumed.\label{fig:lambar}}
\end{figure}
\begin{figure}
\includegraphics[width=\linewidth,angle=180]{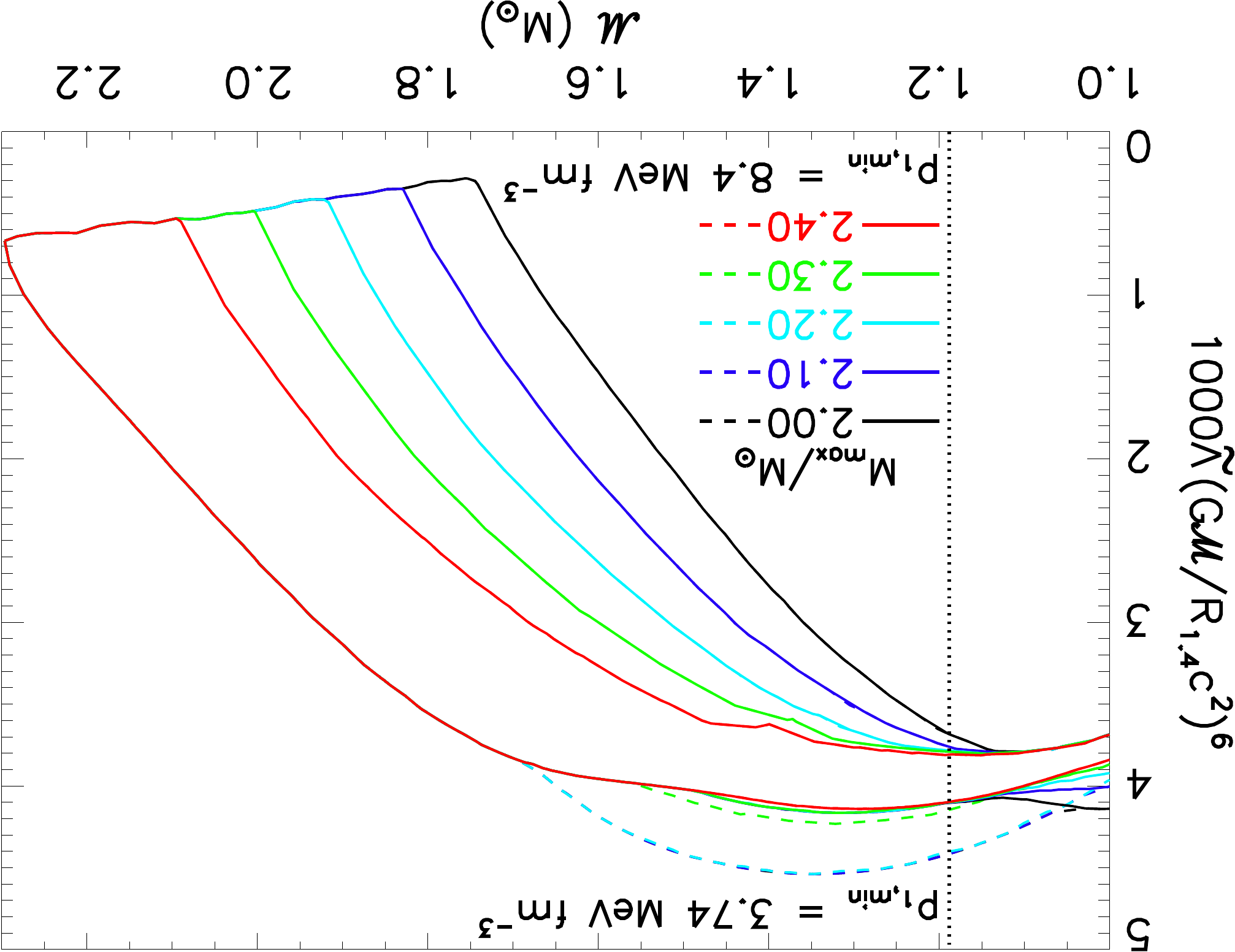}
 \caption{Similar to Fig. \ref{fig:lamb6}, except that bounds of the quantity $\tilde\Lambda[G{\cal M}/(R_{1.4}c^2)]^6$ are displayed. Solid (dashed) lines show bounds for $p_{1.min}=8.4~(3.74)$ MeV fm$^{-3}$.  The chirp mass of GW170817 is shown by the vertical dotted line.\label{fig:lambarb6}} 
\end{figure}
The $\beta$-dependence of $\Lambda$ has interesting consequences for the binary deformability $\tilde\Lambda$, Eq. (\ref{eq:tildelambda}).   For each equation of state in the piecewise polytrope scheme, one can compute $\tilde\Lambda$ for all stellar pairs along the corresponding $M-R$ curve.  The results are displayed in Fig. \ref{fig:lambar}, where equations of state are identified by their corresponding value of $R_{1.4}$, the radius of a $1.4M_\odot$ star.  This figure bears a striking resemblance to Fig. \ref{fig:lam}, and suggests that $\tilde\Lambda\propto({\cal M}/R_{1.4})^{-6}$, at least for values of ${\cal M}\lesssim1.4M_\odot$, a result confirmed in Fig. \ref{fig:lambarb6}.  

As is the case for $\Lambda(M)$, the upper bound of $\tilde\Lambda$ depends on $p_{1,max}$ and is insensitive to a lower limit for $M_{max}$ for ${\cal M}\gtrsim1.1M_\odot$
(Fig. \ref{fig:lambarb6}).  The upper bound for ${\cal M}\lesssim1.6M_\odot$ is  sensitive to $p_{1,min}$.  Similarly, the lower bound to $\tilde\Lambda({\cal M})$ depends both on $p_{1,min}$ and $M_{max}$.

An inferred upper limit to the maximum mass can result in a smaller upper bounds to $\tilde\Lambda$ and $\tilde\Lambda(G{\cal M}/R_{1.4}c^2)^6$, but only for $1.55M_\odot<{\cal M} < M_{max}/2^{1/5}$~\footnote{Note that the maximum chirp mass occurs when $q=1$ and ${\cal M}=M_{max}/2^{1/5}$.}. The maximum reduction to $\tilde\Lambda$ is a factor 2 when $M=M_{max}$.  If $M_{max}=2.5~(\le2.4) M_\odot$ the maximum reduction to $\tilde\Lambda(G{\cal M}/R_{1.4}c^2)^6$ is 0.0002 (0.0005). 

It is interesting to note that Eq. (\ref{eq:lam6}) allows one to express the binary deformability as
\begin{equation}
\tilde\Lambda\simeq{16a\over13}\left({R_{1.4}c^2\over G{\cal M}}\right)^6{q^{18/5}\over(1+q)^{31/5}}\left[r_1^6(1+12q)+r_2^6{12+q\over q^2}\right] ,
\label{eq:tildelambda1}\end{equation}
where $r_i=R_i/R_{1.4}$ and $i$ refers to star $M_1$ or $M_2$.  
For the piecewise polytropes we consider, and in the mass range $1.1M_\odot\le M\le1.6M_\odot$, the radius range is $\Delta R=|R_{M=1.6M_\odot}-R_{M=1.1M_\odot}|\le0.47$  km for all viable equations of state. Moreover, the average spread is only $<\Delta R>\simeq0.1$ km, or less than about 1\%.  Assuming $r_1\simeq r_2\simeq 1$ leads to
\begin{equation}
\tilde\Lambda\simeq{16a\over13}\left({R_{1.4}c^2\over G{\cal M}}\right)^6{q^{8/5}\over(1+q)^{26/5}}(12-11q+12q^2).\label{eq:tildelambda2}
\end{equation}
This equation is remarkably insensitive to $q$.  In fact, one finds
\begin{equation}
\left({\partial\tilde\Lambda\over\partial q}\right)_{{\cal M}}\simeq\tilde\Lambda{(1-q)\over5q(1+q)}\left({96-263q+96q^2\over12-11q+12q^2}\right),
\label{eq:dlambda}\end{equation}
showing the derivative vanishes when $q=1$ and $q=0.434$.  Thus,
$\tilde\Lambda$ is very insensitive to $q$ for the relevant range
$q\gtrsim1/2$ which follows from $M_{2,min}\simeq1M_\odot$ and
$M_{max}\sim2M_\odot$.  In the case of GW170817, $q\gtrsim0.7$ to
90\% confidence~\cite{Abbott17}.  For a given ${\cal M}$, and assuming $r_i=1$, one finds that
$\tilde\Lambda(q=0.7)/\tilde\Lambda(q=1)=1.029$.  Even for $q=0.5$,
the ratio $\tilde\Lambda(q)/\tilde\Lambda(0)=1.11$.  (Indeed, one can
show that $\tilde\Lambda(q=0.274)=\tilde\Lambda(q=1)$.)  Although
$\tilde\Lambda$ is formally a function of ${\cal M}$, $R_1$, $R_2$
and $q$, the effective functional dependence of $\tilde\Lambda({\cal
  M}/R_{1.4})^6$ on $q$ is thus very similar to that of
$\Lambda\beta^6$ on $M$, i.e., 
\begin{equation}
\tilde\Lambda=a^\prime\left({R_{1.4}c^2\over G{\cal M}}\right)^6,
\label{eq:lamb6}
\end{equation}
where $a^\prime=0.0035\pm0.0007$ bounds the results for
$1.0M_\odot\le{\cal M}\le1.4M_\odot$.  However, for GW170817's value
${\cal M}=1.188M_\odot$, one finds $a^\prime=0.0039\pm0.0002$ with
just a $\pm5\%$ variation (Fig. \ref{fig:lambarb6}).   Roughly, $a^\prime$ is determined by setting $q=1$ in Eq. (\ref{eq:tildelambda2}), or $a^\prime\simeq2^{-6/5}a$.  The larger relative range
of $a^\prime$ compared to $a$ is because binaries with ${\cal M}\gtrsim1.2M_\odot$ and small $q$ can contain a massive neutron star $M\gtrsim1.6M_\odot$.

It is useful to invert
Eq. (\ref{eq:lamb6}) to arrive at an estimate for $R_{1.4}$ that is
largely insensitive to the EOS:
\begin{equation}
R_{1.4}\simeq(11.5\pm0.3){{\cal M}\over M_\odot}\left({\tilde\Lambda\over800}\right)^{1/6}{\rm~km}.
\label{eq:r14}
\end{equation}
For GW170817, the accurately known ${\cal M}$ and its inferred $a^\prime$ imply $R_{1.4}\simeq(13.4\pm0.1)(\tilde\Lambda/800)^{1/6}$ km.

\section{Deformability-Mass Correlations for Hadronic Stars\label{sec:hadron}}
An immediate result motivated by the observations with piecewise polytropes that $\Lambda\simeq a\beta^{-6}$ and $r_1\simeq r_2$ is 
\begin{equation}
\Lambda_1\simeq q^6\Lambda_2.
\label{eq:l1l2}\end{equation}
DFLB$^3$ used this correlation in the analysis of the gravitational wave signal from GW170817, allowing a reduction in the number of fitting parameters by one.  Use of this correlation resulted in a better model of the event:  the odds ratio comparing the results including this correlation to not including it was $\gtrsim100$~\citep{De18}.  However, this correlation is not perfect, first because there is a bounding range to $a$ and second, because $dR/dM\ne0$ in the relevant mass range. We now quantify this uncertainty. 

\begin{figure}
\includegraphics[width=\linewidth,angle=180]{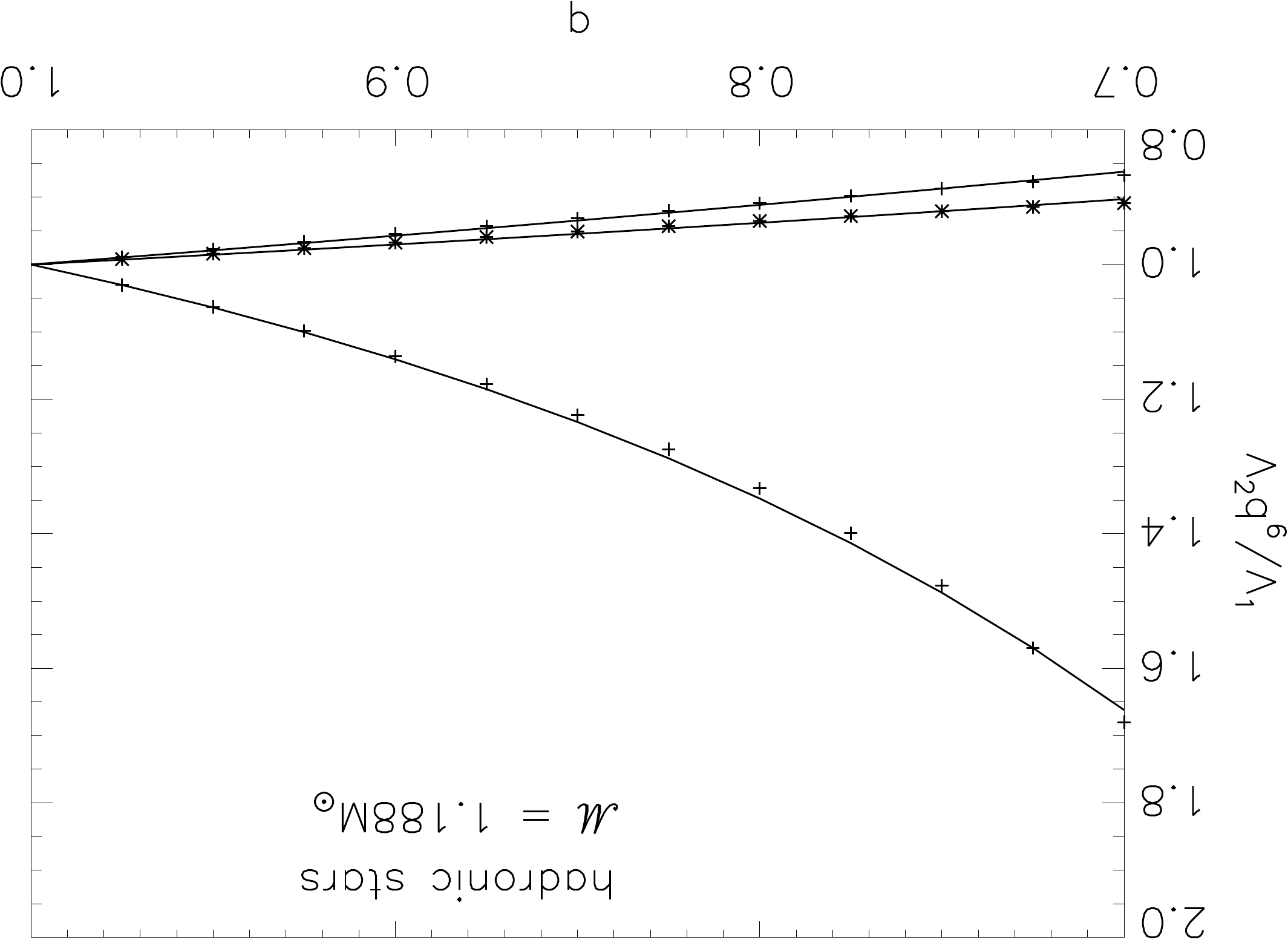}
\caption{Symbols show the upper and lower bounds on
  $\Lambda_2q^6/\Lambda_1$ as a function of $q$ for hadronic stars as determined from piecewise polytropes 
  assuming ${\cal M}=1.188M_\odot$ for GW170817.  The two lower bounds correspond to lower limits $p_{1,min}=3.74$ MeV fm$^{-3}$ (crosses) and 8.4 MeV fm$^{-3}$ (asterisks).  The
  approximate bounds given by Eq. (\ref{eq:llq}) are shown as black
  curves. \label{fig:l2l1q6}}
\end{figure}

\begin{table}[]
    \centering
    \begin{ruledtabular}
    \begin{tabular}{c|cccc}
$p_{1,min}$&3.74 &8.4&\hspace*{-1.5cm}MeV fm$^{-3}$ & \\
\hline\\[-8pt]
${\cal M} (M_\odot)$&$n_-$&$n_-$&$n_{0+}$&$n_{1+}$\\
\hline\\[-8pt]
1.00&5.1717&5.3242&6.4658&-0.24890\\
1.05&5.2720&5.4167&6.7470&-0.32672\\
1.10&5.3786&5.5169&7.0984&-0.44315\\
1.15&5.4924&5.6252&7.5546&-0.62431\\
1.188&5.5839&5.7133&8.0322&-0.86884\\
1.20&5.6138&5.7423&8.1702&-0.91294\\
1.25&5.7449&5.8693&8.9715&-1.3177\\
1.30&5.8960&6.0070&9.9713&-1.8091\\
1.35&6.0785&6.1574&11.234&-2.3970\\
1.40&6.3047&6.3223&12.833&-3.0232\\[-2pt]
    \end{tabular}
    \caption{Hadronic $\Lambda_1/\Lambda_2$ exponents in Eq. (\ref{eq:llq}).}\label{tab:llq}
    \label{tab:my_label}
    \end{ruledtabular}
\end{table}

To begin, for piecewise polytropes, we show upper and lower bounds on
$\Lambda_2 q^6/\Lambda_1$ in Fig. \ref{fig:l2l1q6} that would apply
for GW170817 for which ${\cal M}=1.188M_\odot$ is assumed.  One
observes a spread around the value of unity predicted by
Eq. (\ref{eq:l1l2}) which expands as $q$ decreases.  The lower bound
is determined by the assumed lower limit to $p_1$,
$p_{1,min}$. because those $M-R$ curves can have the largest values of
$(c^2/G)dR/dM$ and hence the smallest ratios of $\Lambda_2/\Lambda_1$
for a given $q$.  We show bounds for the cases $p_{1,min}=3.74$ MeV
fm$^{-3}$, the conservative lower limit from the unitary gas
constraint, and for 8.4 MeV fm$^{-3}$ from neutron matter theoretical
calculations.  On the other hand, the upper limit is determined by the
$M-R$ curves with the minimum possible value of $p_2$, which increases
with the assumed minimum value of the maximum mass
$M_{max}\ge2M_\odot$~\citep{Lattimer16}, because those $M-R$ curves can
have the smallest (i.e., most negative) values of
$(c^2/G)dR/dM$.  
Importantly, we found that  the upper bound to $\Lambda_2/\Lambda_1$, being a ratio, is not sensitive to
$p_{1,max}$ despite the fact that the upper bound to $\Lambda(M)$ is determined by $p_{1,max}$.  We have determined that these
bounds may be approximated as
\begin{equation}
q^{n_-}\ge\Lambda_1/\Lambda_2\ge q^{n_{0+}+qn_{1+}},
\label{eq:llq}
\end{equation}
valid for $q\gtrsim0.65$, where  values for the exponent $n_-$, for the cases that $p_{1,min}=[3.74,8.4]$ MeV fm$^{-3}$, and the exponents $n_{0+}$ and $n_{1+}$ are given in Table \ref{tab:llq}.

\begin{figure}
\includegraphics[width=\linewidth,angle=180]{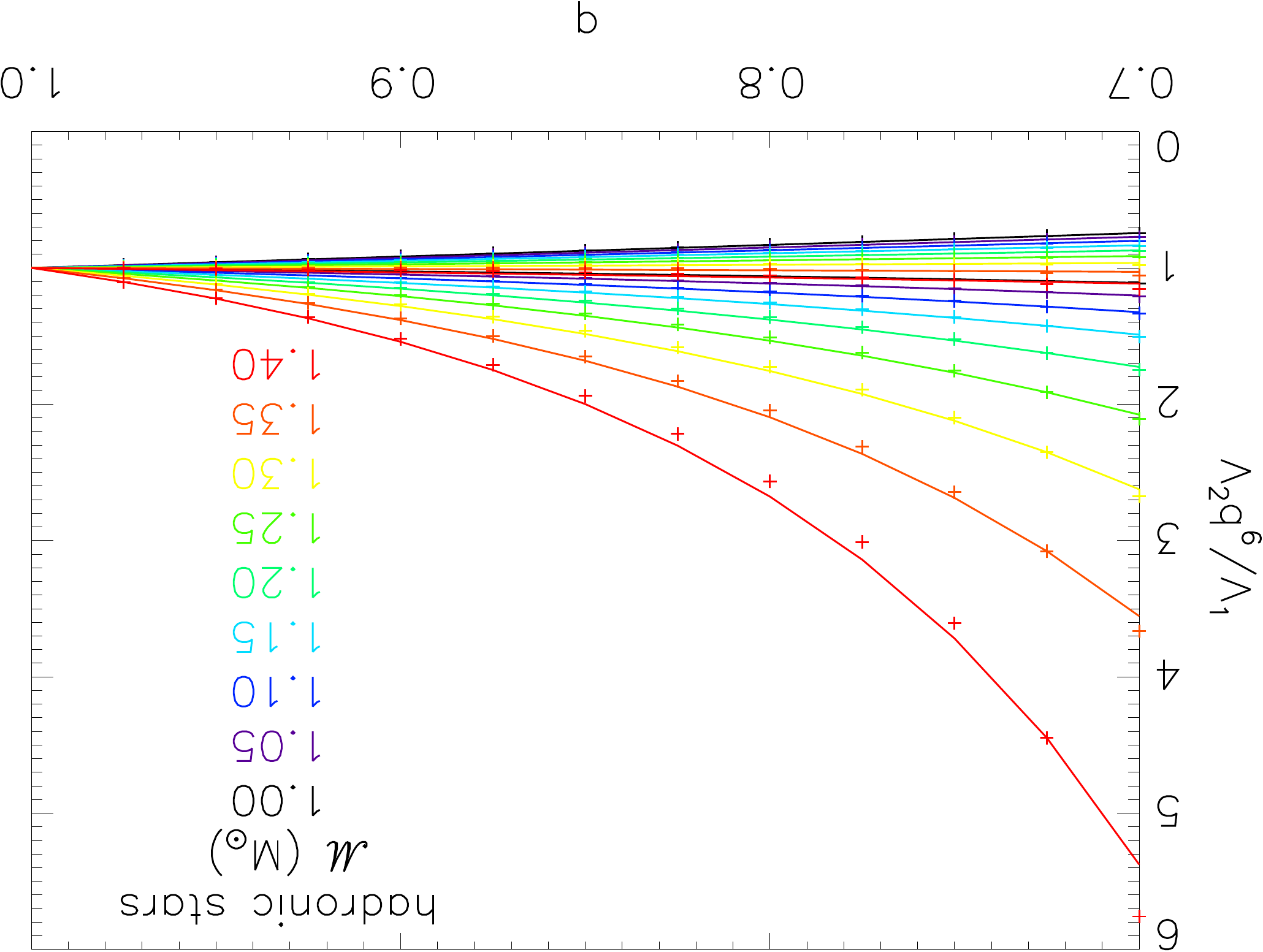}
\caption{The same as Fig. \ref{fig:l2l1q6} but for general chirp mass ranges (color) for hadronic stars.  For clarity, lower bounds using $p_{1,min}=3.74$ MeV fm$^{-3}$ are not shown. \label{fig:l2l1q6g}}
\end{figure}
In future BNS merger events, the chirp masses will likely always be measured to better than $0.01M_\odot$ precision.  It is therefore useful to generalize results to different chirp masses by modifying the exponents. We show results for ${\cal M}$ in the range
$1.0M_\odot\le{\cal M}\le1.4M_\odot$ likely to span future 
mergers in Fig. \ref{fig:l2l1q6g} and summarize the exponents in Table
\ref{tab:llq}.  Bounds for intermediate values can be 
interpolated.  As
before, the lower limits to $\Lambda_2q^6/\Lambda_1$ are determined by
$p_{1,min}$, so the cases $p_{1,min}=3.74$ MeV fm$^{-3}$ and 8.4 MeV
fm$^{-3}$ are shown separately in Table \ref{tab:llq}.  However, they
are so similar they cannot be distinguished on the scale of
Fig. \ref{fig:l2l1q6g}.  The upper limit is determined by $M_{max}$,
which is chosen to be $\ge2M_\odot$; as before, it is not sensitive to
the value of $p_{1,max}$.

\begin{figure*}
\includegraphics[width=\linewidth,angle=0]{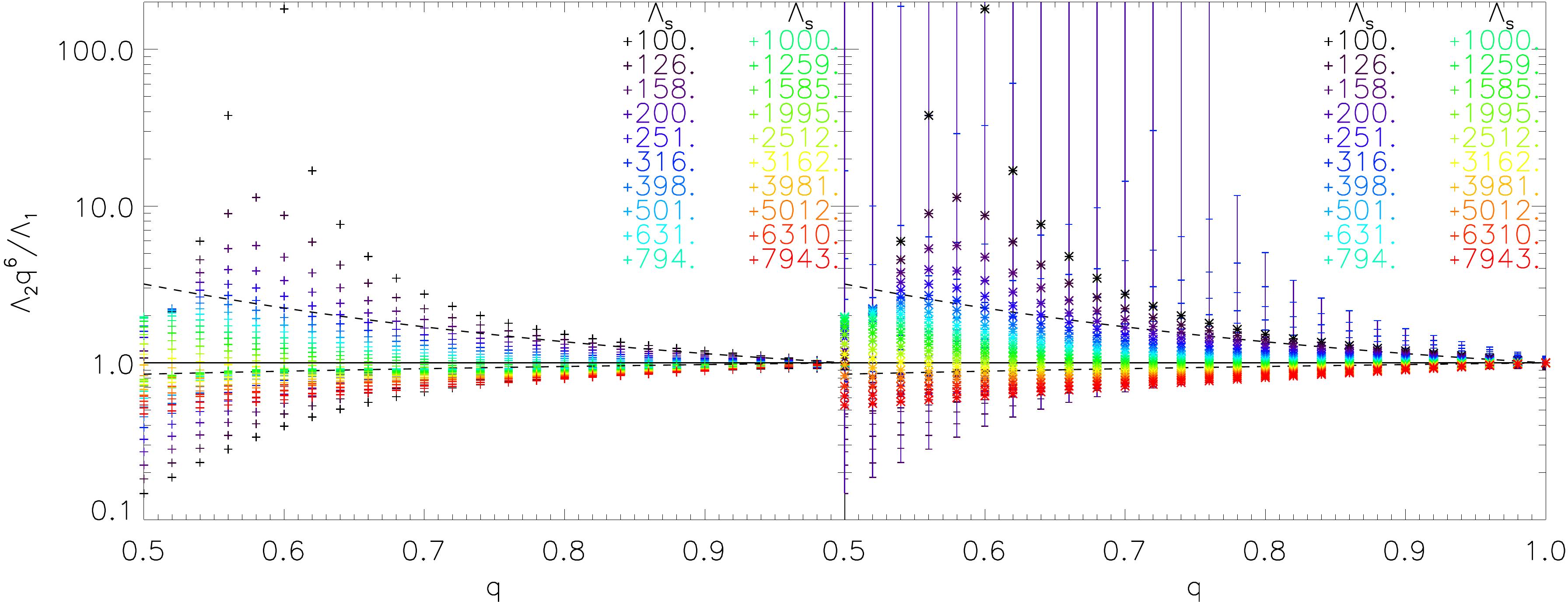}
\caption{The same as Fig. \ref{fig:l2l1q6} but showing the deformability-mass correlation predicted by Ref.~\cite{Yagi17,Chatziioannou18} over all chirp masses. The upper and lower bounds from Eq. (\ref{eq:llq}) for the GW170817 chirp mass of $1.188M_\odot$ are indicated as dashed lines.  The left panel shows the mean value of the quantity $\Lambda_2q^6/\Lambda_1$ as a function of $q$ and $\Lambda_s=\Lambda_1+\Lambda_2$ (indicated by color).  The right panel show mean values as asterisks and their estimated $\pm1\sigma$ uncertainty ranges. \label{fig:comp}}
\end{figure*}

We found that imposing an upper limit to $M_{max}$ does not affect the upper bounds but may slightly increase the lower bounds if $M_{max}<2.2M_\odot$ and $p_{1,min}=3.74$ MeV fm$^{-3}$.  

Another approach was considered by LVC2, who adopted the methodology
of Ref.~\cite{Yagi17}, who fitted 11 realistic equations of state
to determine the optimum value of $\Lambda_2$ as a function of
$\Lambda_1, M_1$ and $M_2$.  They expressed their results in terms of
the symmetric and antisymmetric combinations of $\Lambda_1$ and
$\Lambda_2$: $\Lambda_s=(\Lambda_1+\Lambda_2)/2$ and
$\Lambda_a=(\Lambda_2-\Lambda_1)/2$.  Specifically, they determined an
analytical expression for the optimum fit of $\Lambda_a(\Lambda_s,q)$
which is valid for physically reasonable values of ${\cal M}$.
Ref. \cite{Chatziioannou18} furthermore determined the associated standard
deviations $\sigma_{\Lambda_a}$ for this fit.  For their waveform
modeling, the LVC2 strategy is to sample prior distributions of
$\Lambda_s$ and $q$ values and to then compute associated ranges of
$\Lambda_a$ values, assumed to have a Gaussian distribution with the
aforementioned standard deviations associated with specific choices of
$\Lambda_s$ and $q$.  ${\cal M}$ does not appear
as a specific parameter.  However, this procedure has two disadvantages:
it does not allow sampling of the entire physically-allowed
$\Lambda_a-\Lambda_s$ space, and, in the case of small values of
$\Lambda_s$ and $q$, values of $\Lambda_a>\Lambda_s$ can be selected,
leading to negative values of $\Lambda_1$ and an essentially
unlimited range of $\Lambda_2$ values.

We compare the $1\sigma$ predicted width for
$\Lambda_2q^6/\Lambda_1$ of this procedure with ours for ${\cal
  M}=1.188M_\odot$ appropriate for modeling GW170817 in
Fig. \ref{fig:comp}.  We note that at every $q$, this procedure leads
to a much larger uncertainty range than the bounds we have
established, even without including the $1\sigma$ uncertainty
estimated by Ref.~\cite{Chatziioannou18}.  As mentioned, assuming a Normal distribution with
these uncertainties can lead to the unphysical result that
$\Lambda_2<\Lambda_1$ which has to be excluded.  One reason for the
broader uncertainties with this procedure is that it is not chirp
mass-specific; our results also predict a larger uncertainty range for
larger chirp masses than for the case of GW170817.  This comparison
shows the importance of utilizing information concerning
${\cal M}$, which will be very well determined in a BNS merger, in
modeling deformability-mass correlations.

\section{Minimum Deformabilities From Causality\label{sec:caus}}

It is of interest to determine the correlations among the deformabilities and masses involved in the merger of self-bound stars.  These objects have large finite surface density $\varepsilon_o$ where the pressure vanishes. The idealized case is  a model containing two parameters, $\varepsilon_o$ and a constant sound speed $c_s^2/c^2\equiv s$ for $\varepsilon\ge\varepsilon_o$.  Therefore, the equation of state is simply
\begin{equation}
p=s(\varepsilon-\varepsilon_o); \qquad\varepsilon\ge\varepsilon_o
\label{eq:sb}
\end{equation}
and $p=0$ otherwise.  Koranda, Stergioulas and Friedman~\cite{Koranda97} have conjectured that the most compact stellar configurations, for a given mass $M$, are achieved for the case with $s=1$.  Although not proven, it has been empirically demonstrated that no causal equation of state can produce more compact configurations (see, e.g., Ref.~\cite{Glendenning00}).   This is known as the 'maximally compact' case.   Although there is abundant evidence that observed neutron stars have extensive crusts, largely stemming from observations of pulsar glitches~\citep{Link99,Mongiovi18} and neutron star cooling following transient accretion events~\citep{Ootes18,Chaikin18} and also on longer timescales~\citep{Page04,Beznogov16}, there is no proof that self-bound stars do not, in fact, exist.   

A famous example is the conjecture \citep{Ivanenko65,Witten84,Farhi84} that strange quark matter is the ultimate ground state at zero pressure.  If true, the compression of neutron star cores to sufficiently high density could trigger a phase transition in which most of the hadronic matter is converted to strange quark matter which would be more stable.  Although the detailed equation of state of self-bound strange quark matter is unknown, the essential aspects of their structure can be determined by In the case of the MIT bag model of strange quark matter, the bag constant $B$ is equivalent to $\varepsilon_o/4$ and $s=1/3$.  The equation of state is
$\varepsilon=4B+p/s$, and in order that the strange quark matter have a lower energy per baryon than iron at zero pressure, $E_0<930.4$ MeV, and therefore be more stable than baryonic matter, one requires $B<37.22$ MeV fm$^{-3}$.

For a given value of $s$, Eq. (\ref{eq:sb}) has but a single parameter, $\varepsilon_o$ and so the TOV equations scale with respect to this parameter.  $\varepsilon$, $m$ and $r$ can be replaced by dimensionless variables, i.e.. 
\begin{equation}
w=\varepsilon/\varepsilon_o,\qquad x=r\sqrt{G\varepsilon_o}/c^2,\qquad y=m\sqrt{G^3\varepsilon_o}/c^4.
\label{eq:scale}
\end{equation}
The resulting dimensionless TOV equation can be solved for a family of
solutions determined by the central density, or $w_0=w(x=0)>1$, each
having surface values of radius $x_s(w_0)$ and mass $y_s(w_0)$ that
vary with $w_0$; the surface is where the pressure vanishes, or
$w(x_s)=1$.  Stable solutions exist for $1<w_0<w_{max}$, where
$w_{max}$ is the dimensionless central density of the maximum mass
configuration, i.e., $y_s(w_0)\le y_s(w_{max})$.

\begin{figure}
\includegraphics[width=\linewidth,angle=180]{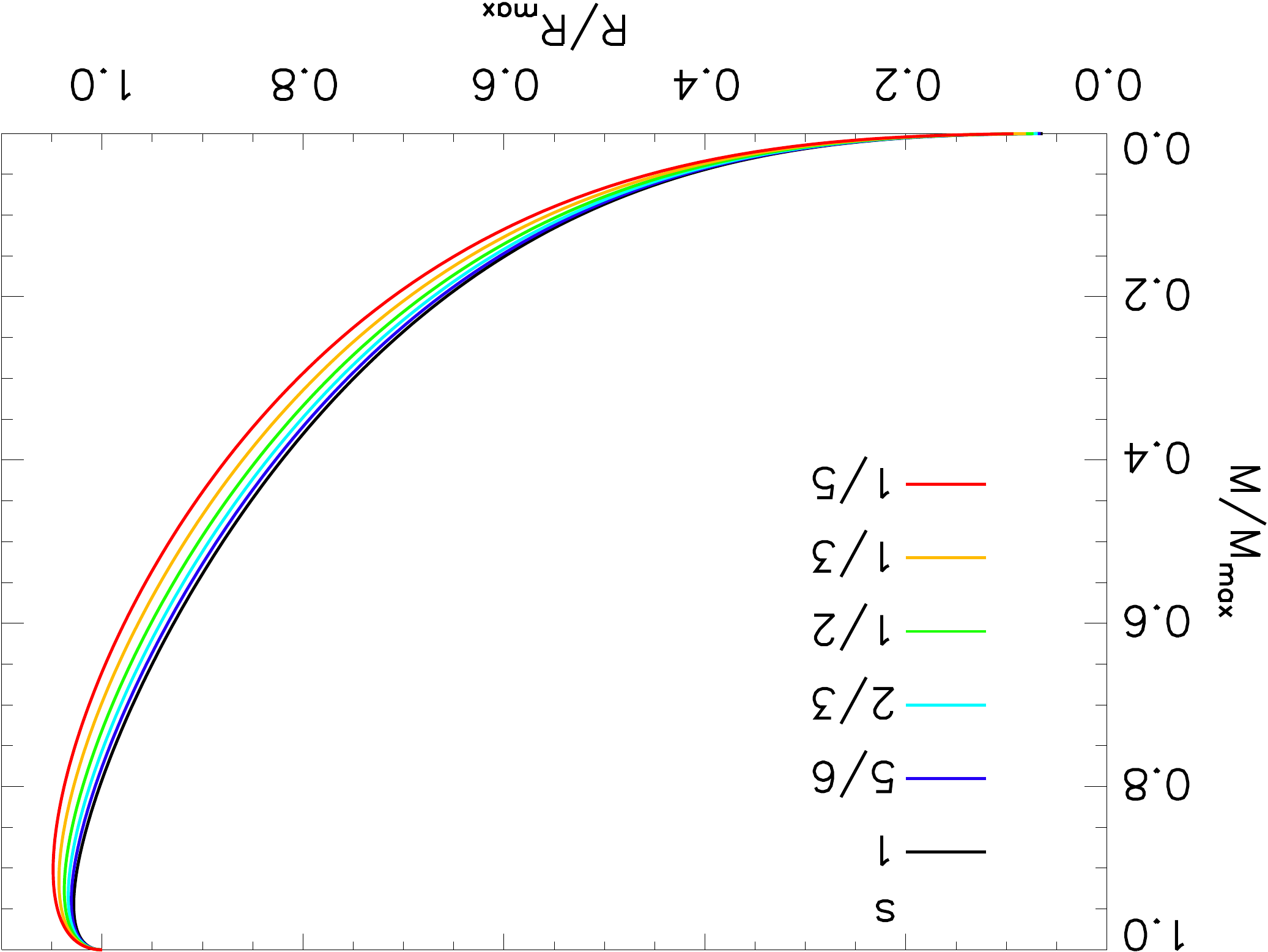}
\caption{The mass-radius curves for self-bound configurations parameterized by the sound speed squared, $s$.  Quantities are normalized relative to their values for the maximum mass solution.\label{fig:mrmax}}
\end{figure}
The solution for which $w_0=w_{max}$ in the case $s=1$ is termed the
maximally compact solution, for which $w_{max}=3.029$,
$x_{s,max}=x(w_{max})=0.2405$ and
$y_{s,max}=y(w_{max})=0.08513$~\citep{Lattimer11}.  The resulting
$M-R$ relation, parametrically expressed as $y(w_0)-x(w_0)$ for
$1<w_0<w_{max}$, has the smallest radius for a given mass for any
causal equation of state in general relativity.  The largest value of
$\beta=GM/(Rc^2)=y_s/x_s$ is
$\beta_{max}=y_{s,max}/x_{s,max}=0.3542=1/2.824$, but less
compact configurations are also excluded for masses smaller than the
maximum mass.  By employing the mass of the most massive
accurately-measured pulsar, $M_{max}=2.01\pm0.04M_\odot$~\citep{Antoniadis13}, one can then
determine the most compact $M-R$ boundary from the parametric
equations
\begin{eqnarray}
M&=&M_{max}{y_s(w_0)\over y_{s,max}}, \cr
R&=&R_{max}{x_s(w_0)\over x_{s,max}}={GM_{max}\over c^2}{x_s(w_0)\over y_{s,max}}.
\label{eq:MRmax}
\end{eqnarray}
$R_{max}$ is the radius of the maximum mass solution.
As $M_{max}$ is increased, the minimum causal radius is increased for every $M<M_{max}$.  Fig. \ref{fig:mrmax} shows the maximally-compact solution in the dimensionless variables $M/M_{max}=y_s(w_0)/y_s(w_{max})$ and $R/R_{max}=x_s(w_0)/x_s(w_{max})$.  Since $M_{max}$ is currently $\simeq2M_\odot$, this figure is easy to interpret in terms of solar masses and km (for $s=1$, $M_{max}=2M_\odot$ corresponds to $R_{max}=8.34$ km).  Similar mass-radius curves and maximum compactnesses $\beta_{max}$ for other values of $s$ are displayed in Fig. \ref{fig:mrmax} and Table \ref{tab:lamminfit}, respectively.

\begin{table}
\begin{ruledtabular}
\begin{tabular}{lllllll}
$s$&1&5/6&2/3&1/2&1/3&1/5\\
\hline \\[-8pt]
$w_{max}$&3.029&3.2404&3.544&4.008&4.816&6.095\\
$x_{s,max}$&0.2405&0.2331&0.2235&0.2104&0.1909&0.1652\\
$y_{s,max}$&0.08513&0.07992&0.07328&0.06439&0.05169&0.03648\\
$\beta_{max}$&0.3542&0.3429&0.3279&0.3060&0.2708&0.2209\\
\hline \\[-8pt]
$a_0$&13.42&13.61&13.91&14.31&15.04&16.15\\
$a_1$&-23.04&-22.82&-22.71&-22.39&-22.11&-21.54\\
$a_2$&20.56&20.32&20.27&19.92&19.71&19.10\\
$a_3$&-9.615&-9.461&-9.398&-9.174&-9.005&-8.639\\[-2pt]
\end{tabular}
\end{ruledtabular}
\caption{Maximally-compact EOS maximum mass solutions and fitting coefficients for Eq. (\ref{eq:lamminfit}).\label{tab:lamminfit}}
\end{table}
\begin{figure}
\includegraphics[width=\linewidth,angle=180]{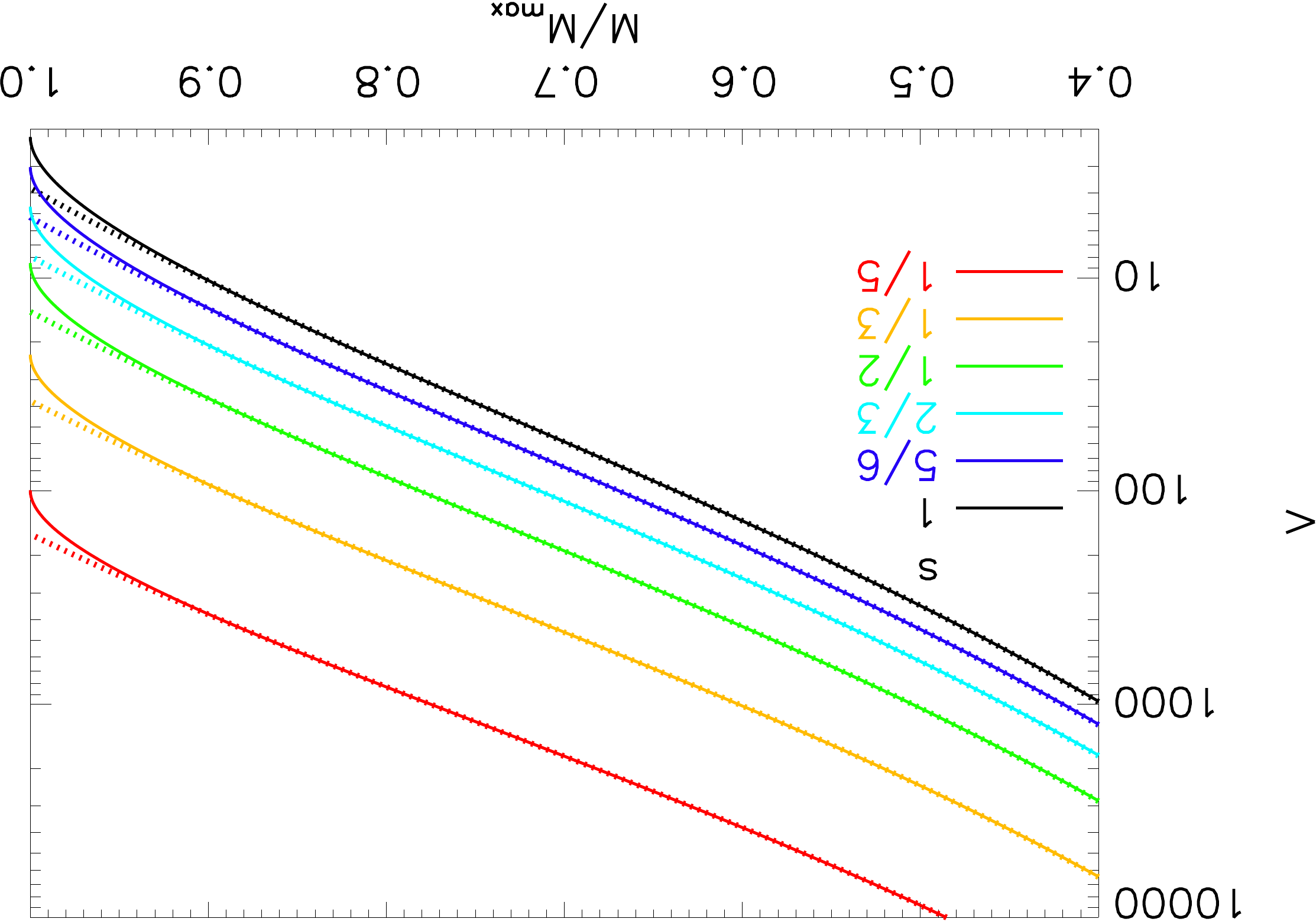}
\caption{The dimensionless deformability as a function of $M/M_{max}=y(w_0)/y_{s,max}$ for self-bound stars parameterized by a constant sound speed $c_s^2/c^2=s$.  Dotted curves show cubic polynomial fits using Eq. (\ref{eq:lamminfit}).\label{fig:lammin}}
\end{figure}
One may now solve  Eq. (\ref{eq:z}) determining the tidal deformability. The variable $z$ is already dimensionless and does not need to be rescaled, but snce  a density discontinuity exists at the surface, the correction described in Sec. \ref{sec:tide} must be applied.   Fig. \ref{fig:lammin} shows the dimensionless deformability $\Lambda$ as a function of $M/M_{max}$ for the maximally compact solution $s=1$.  For the specific case that $M=1.4M_\odot$ and $M_{max}=2M_\odot$, one can see that $\Lambda(1.4M_\odot)\simeq59$.  By conjecture, this currently is the causal minimum value of the deformability for a $1.4M_\odot$ star, but its value will increase by a factor $\simeq(M_{max}/2.0M_\odot)^{5.5}$ if $M_{max}$ is increased.  Similar deformability-mass curves may be computed for other values of $s$ (Fig. \ref{fig:lammin}).  For $0.3\lesssim M/M_{max}\lesssim0.95$, these results may be approximated with cubic polynomials whose coefficients are given in Table \ref{tab:lamminfit}:
\begin{equation}
\ln\Lambda=\sum_{i=0}^3a_i\left({M\over M_{max}}\right)^i
\label{eq:lamminfit}
\end{equation}

\begin{figure}
\includegraphics[height=\linewidth,angle=90]{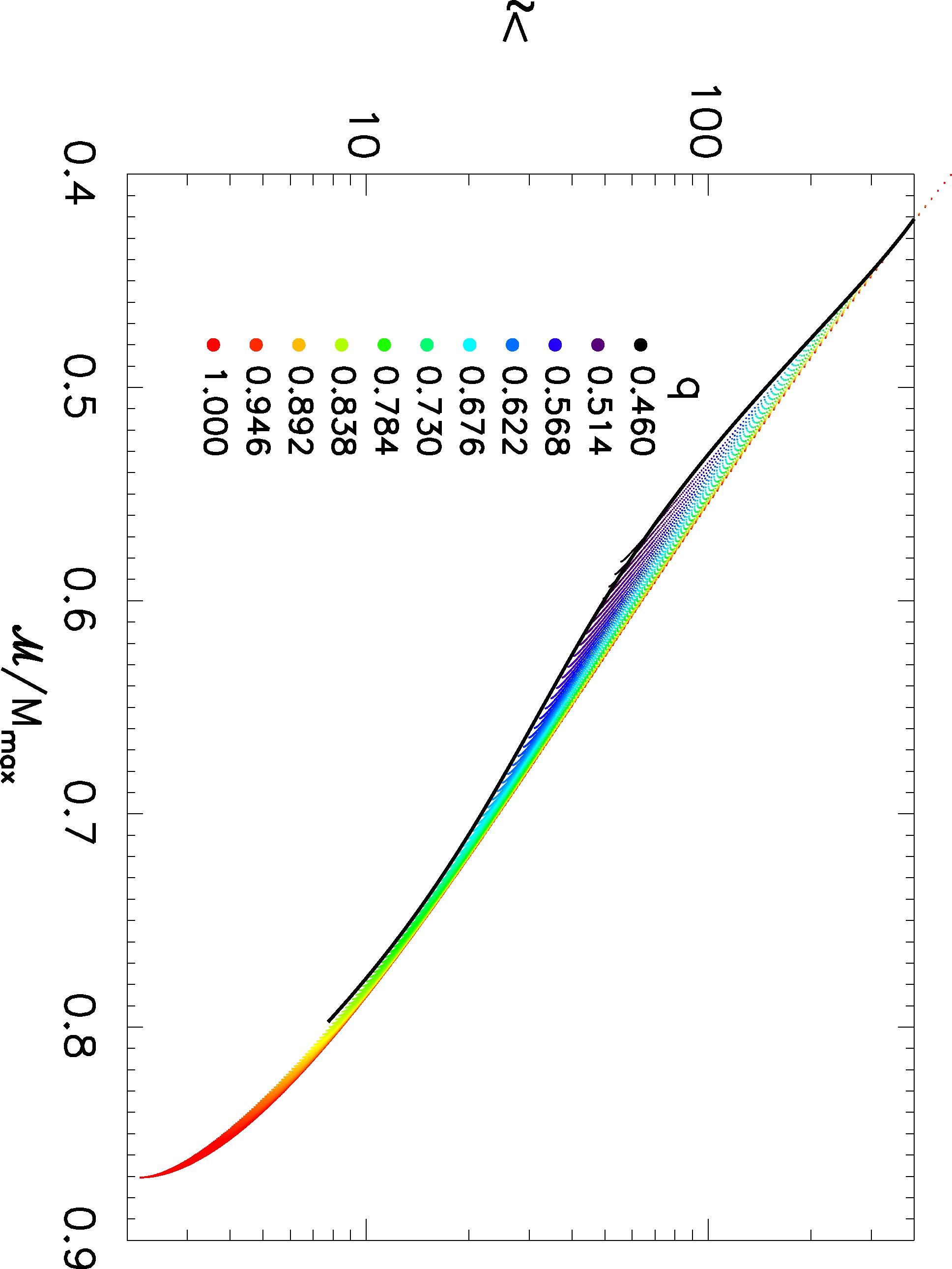}
\caption{The binary  deformability as a function of ${\cal M}/M_{max}$ for the maximally compact self-bound stars with $s=1$.  Binary pairs are shown by points color coded according to their mass ratio $q$.  The solid curve is a quintic polynomial approximation for the lower boundary using Eq. (\ref{eq:lambsb}).\label{fig:lambsb}}
\end{figure}
Because we can give $\Lambda(M)$ explicitly for self-bound stars,
computing $\Lambda_1/\Lambda_2$ as a function
of $q$ and ${\cal M}$ is trivial.  It is also straightforward to determine the binary deformability $\tilde\Lambda$ of self-bound stars.  The results again scale with the assumed value of $M_{max}$ and are shown in Fig. \ref{fig:lambsb} for $s=1$.  By conjecture, these are the minimum  causally-allowed  binary deformabilities for any binary.  The lower boundary can be fit with
\begin{multline}
\tilde\Lambda_{min}\simeq-244.86z^{-6}+2058z^{-5}-6723.2z^{-4}+\\
  + 10760z^{-3}-8428.3z^{-2}  +2582.5z^{-1},
\label{eq:lambsb} 
\end{multline}
where $z={\cal M}/M_{max}$, for $0.45<z<0.8$.
This is therefore the causal minimum for $\tilde\Lambda({\cal M}/M_{max})$. For the case of GW170817, ${\cal M}=1.188M_\odot$, so if $M_{max}\ge2M_\odot$, one sees that $\tilde\Lambda_{min}\ge51$.  Note that using Eq. (\ref{eq:r14}) one then obtains $R_{1.4}\ge8.43$ km whereas the exact causal minimum with $M_{max}=2M_\odot$ is 8.34 km, demonstrating the validity of this equation even beyond the ranges expected for hadronic stars.

\section{Minimum Deformabilities from the Unitary Gas and Neutron Matter Constraints\label{sec:unitary}}
Tews et al.~\cite{Tews17} argue that a robust lower limit to the energy of neutron matter, and therefore effectively that of neutron star matter above the nuclei-gas phase transition around $n_s/2$, is the energy of an idealized unitary gas, which is
\begin{equation}
E_{UG}=\xi_0E_{FG}={3\xi_0\over5}{\hbar^2\over2m}(3\pi^2n_su)^{2/3},
\label{eq:ug}
\end{equation}
where $E_{FG}$ is the energy of a non-interacting Fermi gas, $u=n/n_s$, and $\xi_0\simeq0.37$ is the experimentally-measured Bertsch constant.
 If true, this automatically sets a lower limit to the neutron pressure $p_{N}$:
\begin{equation}
p_{N}\ge n_su^2{\partial E_{UG}\over\partial u}=\xi_0n_s{\hbar^2\over5m}(3\pi^2n_s)^{2/3}u^{5/3}.
\label{eq:pg}
\end{equation}
Assuming that the neutron star matter pressure is approximately equal to the neutron pressure, at the density $n_1=1.85n_s$ we find 
$p_1\ge3.74{\rm~MeV~fm}^{-3}$.  On the other hand, theoretical calculations of the properties of neutron matter~\citep{Drischler16} give appreciably larger values at this density, $p_1\gtrsim8.4$ MeV fm$^{-3}$ as we utilized in \S \ref{sec:hadron}.

In the unitary gas limiting case where $p_{1,min}=3.74$ MeV fm$^{-3}$, the energy Eq. (\ref{eq:ug}) cannot be used to arbitrarily large densities because the $2M_\odot$ maximum mass constraint would be impossible to satisfy.  However, for hadronic stars, one could use this energy up to the density $n_1$ and then, subject to causality, arbitrarily increase the energy at higher densities to ensure compliance with $M_{max}$.  This situation can be approximated by setting $p_{1,min}=3.74$ MeV fm$^{-3}$ and employing the piecewise polytrope scheme as before.  The lower bound to radii will once again be determined by the assumed value of $M_{max}$, but will be smaller than shown in Fig. \ref{fig:mvsr}.  As previously mentioned, if $M_{max}=1.90M_\odot$, $R_{1.4}$ can be as small as 10.5 km.  Similarly, the lower bound to $\Lambda(M)$ will also decrease with $p_{1,min}$ for each value of $M_{max}$.  While $\Lambda_{min}(1.4M_\odot)\simeq197$ in the realistic neutron matter limiting case that $p_{1,min}=8.4$ MeV fm$^{-3}$ and $M_{max}=2M_\odot$ (Fig. \ref{fig:lam}), for $p_{1,min}=3.74$ MeV fm$^{-3}$ (the unitary gas limiting case) and the same $M_{max}$ it is about 156.  We have fit the lower bounds $\Lambda_{min}(M)$ for both values of $p_{1,min}$, for various values of $M_{max}$, using 
\begin{equation}
\ln\Lambda_{min}=\sum_{i=0}^3b_i(M/M_\odot)^i,
\label{eq:lammin}
\end{equation}
where the coefficients $b_i$ are provided in Table \ref{tab:lammin}.  These fits are valid for $1M_\odot<M<0.95M_{max}$.  
\begin{table}
\begin{ruledtabular}
\begin{tabular}{lcccccc}
$p_{1,min}$&$M_{max}$&2.0&2.1&2.2&2.3&2.4\\
\hline \\[-8pt]
\multirow{4}{1cm}{$3.74$ MeV fm$^{-3}$}&$b_0$&17.329&17.345&16.176&15.047&14.572\\
&$b_1$&-17.947&-17.354&-14.497&-11.902&-10.776\\
&$b_2$&9.8648&9.0022&6.7804&4.8887&4.0766\\
&$b_3$&-2.3640&-2.0178&-1.4319&-0.96710&-0.76617\\
\hline \\[-8pt]
\multirow{4}{1cm}{8.4 MeV fm$^{-3}$}&$b_0$&18.819&17.700&16.572&15.534&15.131\\
&$b_1$&-19.862&-17.191&-14.358&-12.011&-11.708\\
&$b_2$&10.881&8.6973&6.5452&4.8485&4.1825\\
&$b_3$&-2.5713&-1.9458&-1.3822&-0.96191&-0.79197\\[-2pt]
\end{tabular}
\end{ruledtabular}
\caption{Coefficients for $\Lambda_{min}$ fits from Eq. (\ref{eq:lammin}) for hadronic stars for both the unitary gas limit and the realistic neutron matter cases.}\label{tab:lammin}
\end{table}

\section{Deformability-Mass Correlations of Hybrid Stars\label{sec:hybrid}}
We so far have largely ignored the possibility of strong first-order
phase transitions in neutron stars.  An important issue is how much
the correlation between the deformabilities is broadened by the
possible appearance of a different phase of matter, such as deconfined
quark matter, in the relevant density range between the central
densities $n_{c,[1,2]}$ of the two stars.  This could substantially
reduce the value of $R_1$ and thereby break the condition $R_1\simeq
R_2$ even for stars of almost the same mass.  Configurations with such a phase transition are often called
hybrid stars (as opposed to purely hadronic stars), and it is of
interest to determine if gravitational-wave signals could provide
support for or against their existence.  Should the more massive star be a hybrid star, and the lower mass star be a hadonic star, the bounds on $\Lambda_1/\Lambda_2$ will be much larger than if both are hadronic or hybrid stars.  In this paper, we establish
analytic absolute bounds for values of $\Lambda_1/\Lambda_2$ for
hybrid stars subject to similar constraints as assumed for purely
hadronic stars.  The piecewise polytrope methodology adopted does
allow a first order phase transition at the pressure $p_2=p_3$
spanning the interval $n_2\le n\le n_3$; however, this is a serious
restriction to what might be possibles.  We here consider a more
general method of introducing phase transitions that does not require
these restrictions.  We will demonstrate that useful
bounds on this correlation can still be analytically expressed as functions
of $q$ and ${\cal M}$.

To construct families of hybrid stars, we follow the methodology of
Ref.~\cite{Alford13} who model phase transitions with three parameters: the
pressure $p_t$ where they occur, the fractional energy density change
across the transition $\Delta\varepsilon_t/\varepsilon_t$, and the
sound speed of matter $s=c^2_s/c^2$ for the new phase, which is
assumed to be constant, for $p>p_t$.  \cite{Alford13} shows that the
phase space allowed for strong phase transitions increases with $s$,
and for $s\le1/3$ there is almost no phase space allowed for hybrid configurations once the
$M_{max}=2M_\odot$ constraint is considered.  As a result, to consider
the maximum bounds for $\Lambda_2/\Lambda_1$ we focus on the extreme, and
possibly unrealistic, case $s=1$.  We employ the three-segment piecewise polytropic equation of state for hadronic matter with $p\le p_t$, but we allow for phase transitions with $p_t\ge p_s$ and $\Delta\varepsilon_t/\varepsilon_t>0$ limited from above by the maximum mass constraint.

\begin{figure}
\includegraphics[height=\linewidth,angle=90]{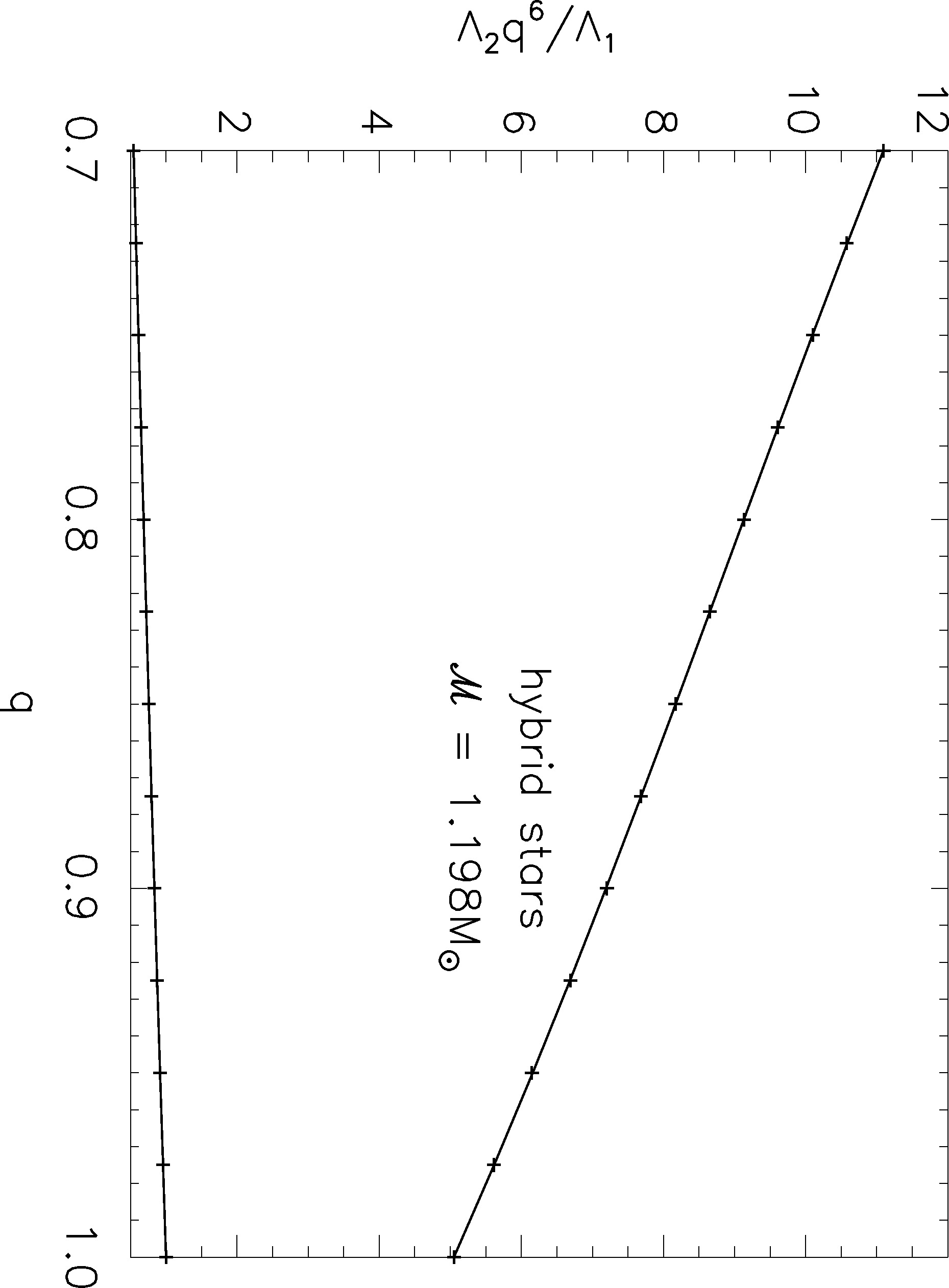}
\caption{Symbols show the upper and lower bounds on
  $\Lambda_2q^6/\Lambda_1$ as a function of $q$ for hybrid stars as determined from piecewise polytropes 
  assuming ${\cal M}=1.188M_\odot$, 
  appropriate to GW170817.  The lower bound corresponds to $p_{1,min}=8.4$ MeV fm$^{-3}$.  The upper bound corresponds to $M_{max}\ge2M_\odot$ and $p_{1,max}=30$ MeV fm$^{-3}$.  The
  approximate bounds given by Eq. (\ref{eq:llqhy}) are shown as black
  curves. \label{fig:l2l1q6hy}}
\end{figure}
\begin{table}
\begin{ruledtabular}
\begin{tabular}{c|cccccc}
\\[-10pt]
$p_{1,min}$&3.74&8.4&\multicolumn{2}{l}{\hspace*{-10pt}MeV fm$^{-3}$}&&\\
\hline\\[-8pt]
${\cal M} (M_\odot)$&$n_-$&$n_-$&$n_{0+}$&$n_{1+}$&$n_{2+}$&$n_{3+}$\\
\hline \\[-8pt]
1.00&4.1555&4.1788&-0.74665&3.3267&-4.4057&1.9998\\
1.05&4.1932&4.2162&-0.95564&4.0789&-5.3424&2.4010\\
1.10&4.2307&4.2524&-1.1902&4.9075&-6.3577&2.8293\\
1.15&4.2707&4.2889&-1.3230&5.3650&-6.9267&3.0792\\
1.188&4.2995&4.3187&-1.4475&5.7829&-7.4254&3.2872\\
1.20&4.3112&4.3281&-1.4160&5.6484&-7.2500&3.2147\\
1.25&4.3502&4.3673&-1.6317&6.3747&-8.1036&3.5580\\
1.30&4.3932&4.4089&-1.8586&7.1188&-8.0499&3.8838\\
1.35&4.4362&4.4517&-1.9485&7.3619&-9.1952&3.9703\\
1.40&4.4808&4.4954&-2.1439&7.9539&-9.8241&4.1954\\[-2pt]
\end{tabular}
\end{ruledtabular}
\caption{Hybrid star $\Lambda_1/\Lambda_2$ parameters in Eq. (\ref{eq:llqhy}).}\label{tab:llqhy}
\end{table}
We first examine the bounds for the case applicable to GW170817, namely ${\cal M}=1.188M_\odot$.  Fig. \ref{fig:l2l1q6hy} displays the upper and lower bounds for $\Lambda_2q^6/\Lambda_1$, assuming $M_{max}\ge2.0M_\odot$ and 3.74 MeV fm$^{-3}\le p_1\le30$ MeV fm$^{-3}$.
The lower bound depends weakly on $p_{1,min}$ (Table \ref{tab:llqhy}), and can be approximately described as $q^{n_-}$ as in the hadronic case (the alternate lower bound from $p_{1,min}=3.74$ MeV fm$^{-3}$ cannot be distinguished in this figure).  We found that imposing an upper limit to $M_{max}$ below about $2.4M_\odot$ can increase the lower bound, but we do not consider that further here.  In contrast to the purely hadronic case, the upper bound depends strongly on $p_{1,max}$, because $R_2$ depends strongly on this but $R_1$ (now a hybrid star) does not.  The upper bound weakly depends on the minimum value of $M_{max}$.  Even for $q\simeq1$, one finds if a strong phase transition occurs at the central density of a star with mass $M_2\simeq M_1$, one has $R_1<R_2$ and $\Lambda_1<\Lambda_2$ since $\Lambda\propto(R/M)^6$.  For hybrid stars, the upper boundary can be approximated with a cubic polynomial $q$-dependence:
\begin{equation}
q^{n_-}\ge\Lambda_1/\Lambda_2\ge\sum_{i=0}^3 n_{i+}q^i,
\label{eq:llqhy}
\end{equation}
where parameter values are given in Table \ref{tab:llqhy}.

\begin{figure}
\includegraphics[width=\linewidth,angle=180]{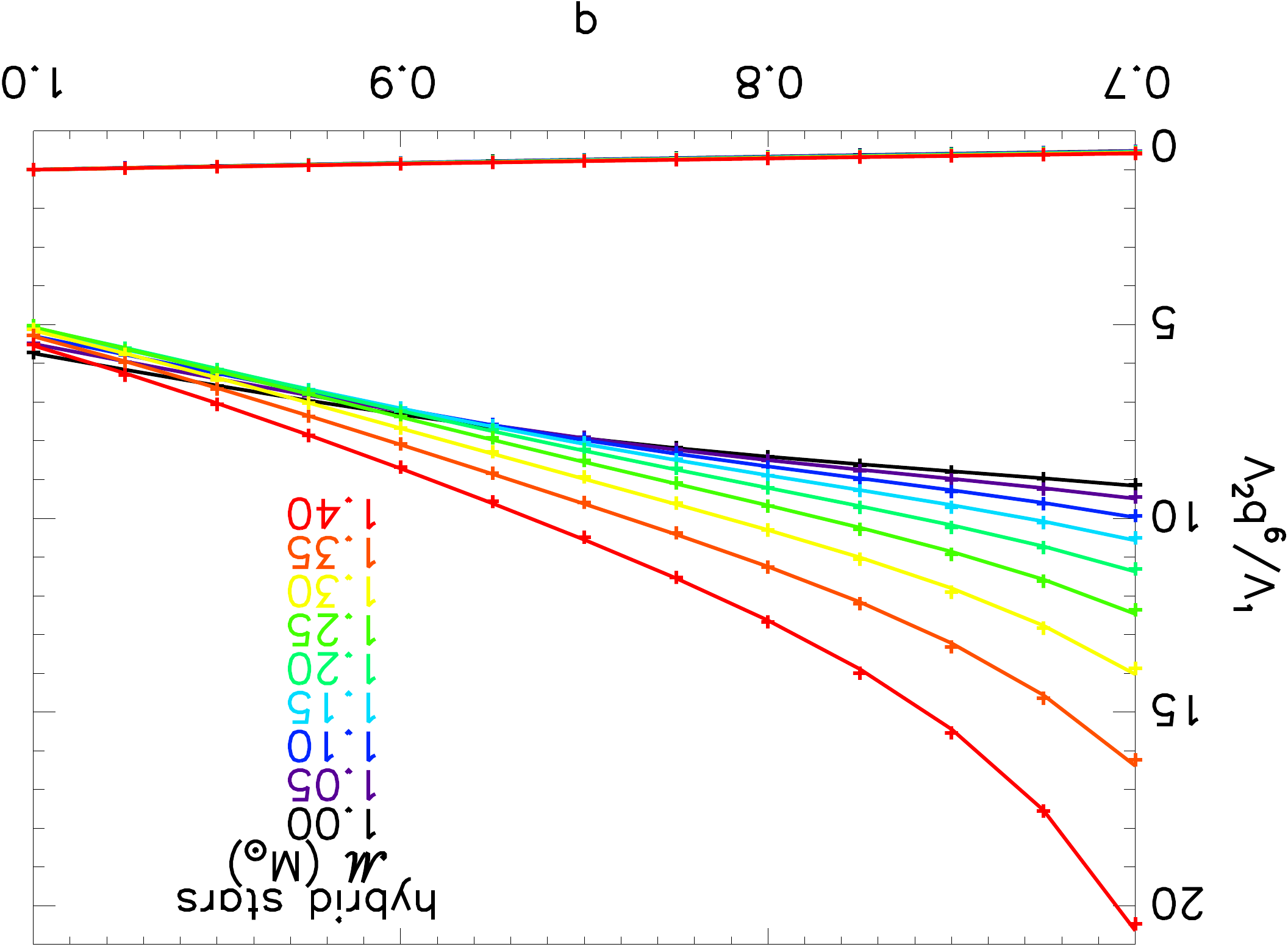}
\caption{The same as Fig. \ref{fig:l2l1q6hy} but for general chirp mass ranges (color) for binaries with one hybrid star.  For clarity, lower bounds using $p_{1,min}=3.74$ MeV fm$^{-3}$ are not shown. \label{fig:l2l1q6ghy}}
\end{figure}

Results for general chirp masses are displayed in Fig. \ref{fig:l2l1q6ghy}; in all cases, as for hadronic stars, the two lower bounds for different values of $p_{1,min}$ cannot be distinguished on the scale of the figure. The lower bounds are also insensitive to ${\cal M}$ because the corresponding configurations are close to the maximally compact ones.   Upper bounds depend, as for the hadronic stars, on $p_{1,max}$, which is chosen to be 30 MeV fm$^{-3}$ for this figure. Coefficients $n_-$ for the lower bound and $n_{i+}$ for the upper bound, using Eq. (\ref{eq:llqhy}), are listed in Table \ref{tab:llqhy}.

Imposing an upper limit to $M_{max}$ does not change the upper bounds to $\Lambda_2/\Lambda_1$ in the hybrid case, but if $M_{max}\lesssim2.6M_\odot$, the lower bounds are increased at $q=0.7$ by up to 10\% (50\%) for ${\cal M}=1 (1.4) M_\odot$, the effect increasing with decreasing $M_{max}$.

Minimum values for $R(M)$ and $\Lambda(M)$ in the case of hybrid stars will be achieved when a phase transition occurs at the smallest possible density that still satisfies the assumed value of $M_{max}$.  We assume that the transition density is no smaller than $n_s$, for which the transition pressure $p_t=p_s$ will depend on $p_{1,min}$ through $p_t=p_0(n_s/n_0)^{\gamma_1}$ where $\gamma_1=\ln(p_{1,min}/p_0)/\ln(n_1/n_0)$ takes the values 1.77 and 2.27 for the cases $p_{1,min}=3.74$ MeV fm$^{-3}$ and 8.4 MeV fm$^{-3}$, respectively.  We find $p_t=p_s=1.18$ MeV fm$^{-3}$ and 1.90 MeV fm$^{-3}$, respectively.  These pressures are so small compared to the central pressures that the effective values of $\Lambda_{min}(M)$ for hybrid stars are the same as $\Lambda(M)$ for the maximally compact EOS for the case $s=1$.

\section{Discussion and Conclusions\label{sec:conclusion}}
In this paper, we have established upper and lower bounds for $\Lambda_2/\Lambda_1$ as functions of $q$ and ${\cal M}$, and minimum values of $\Lambda(M)$, that can be used to restrict the priors of deformabilities in analyses of gravitational-wave data from neutron star mergers.  DFLB$^{3}$ has shown that taking these correlations and bounds into account significantly improves fits in the case of GW170817.  Imposing correlations reduced the uncertainty range for $\tilde\Lambda$, lowering the 90\% credible upper limit by approximately 20\%.  

\begin{figure}
\includegraphics[width=\linewidth,angle=180]{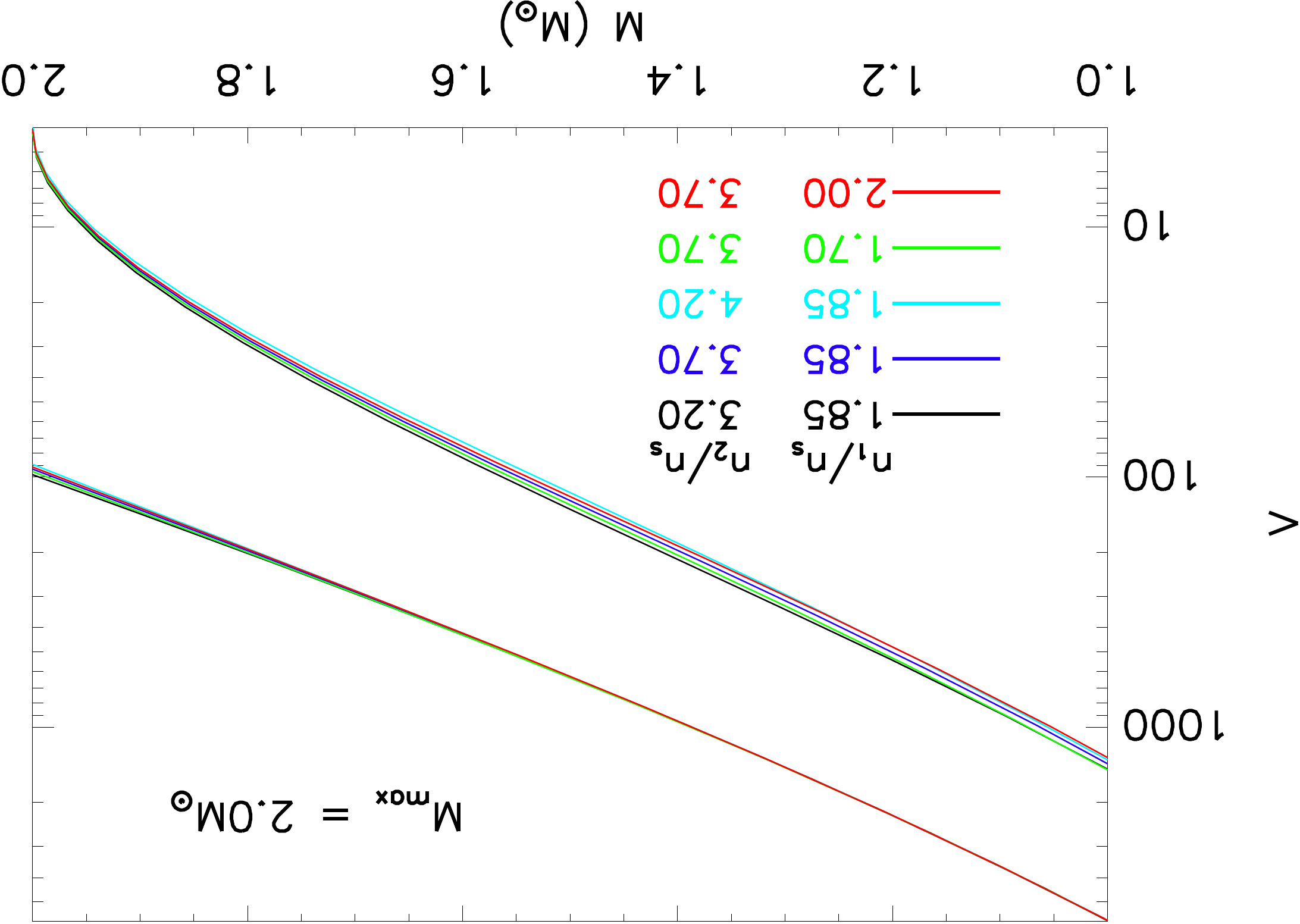}
\caption{The variation of the upper and lower bounds to $\Lambda(M)$ for hadronic stars
  as the boundary densities $n_1$ and $n_2$ are changed.
  $M_{max}=2.0M_\odot$ is assumed.\label{fig:hbound}}
\end{figure}

The bounds we established for hadronic stars were based on a piecewise
polytropic scheme with three segments and fixed boundary densities.
We find our results with three segments to be relatively
insensitive to reasonable variations of the boundary densities
(Fig. \ref{fig:hbound}) $n_1$ and $n_2$.  Varying the boundary
densities produce variations of order $\pm5\%$ in the upper boundary
and $\pm10\%$ in the lower boundary although, for a $1.4M_\odot$ star,
the maximum value of $\Lambda$ is about 6 times the lowest value for
$M_{max}=2M_\odot$.

However, the variations produced by altering the number of polytropic
segments can be more extreme.  Adding polytropic segments allows for
the possibility of one or more strong first-order phase transitions
and so the upper and lower bounds to $\Lambda(M)$ can approach the
results for the hybrid configurations in these cases.
However, restricted to parameter ranges that approximate purely hadronic equations of state, 
varying the number of polytropic segments produce changes to 
$\Lambda(M)$ bounds similar to the changes induced by altering the
boundary densities in the three-polytrope scheme shown in
Fig. \ref{fig:hbound}.

Modifying the piecewise polytrope scheme to smooth its behavior near the segment boundaries, as in the spectral decomposition method \cite{Lindblom18}, also has been shown to increase the accuracy in reproducing specific equations of state. Other high-density approximation methods have also been suggested, e.g., Ref.~\cite{Kurkela14}. However, such schemes inevitably reduce the allowed ranges of sampled pressure-density relations and therefore result in artificially smaller bounding ranges.  It is important to emphasize that determining $\Lambda(M)$ bounds is dissociated from the question of a parameterized scheme's accuracy in reproducing $\Lambda(M)$ from a specific equation of state.  Nevertheless, if one attempts to directly deduce the EOS itself from gravitational waveform modeling, as LVC2 has attempted, the accuracy of the high-density approximation scheme becomes an important consideration. 
\section*{Acknowledgements}
This work was stimulated by the KITP Rapid Response Workshop: Astrophysics from a Neutron Star Merger, and by the INT Program INT-18-72R: First Multi-Messenger Observations of a Neutron Star Merger and its Implications for Nuclear Physics.
JML thanks the hospitality of the KITP and the INT.  We acknowledge fruitful discussions with F. Douglas Swesty, Soumi De, Duncan Brown, B. Sathyaprakash, Samaya Nissanke, Tanja Hinderer and Sophia Han.  
This work was supported in part by
US DOE Grant DE-AC02-87ER40317 and NASA Grant 80NSSC17K0554.


\end{document}